\documentclass[11pt]{article} \textheight=22 true cm
\usepackage{graphicx} 
\usepackage{pslatex}
\usepackage{subfigure}
\usepackage{caption}
\usepackage{subcaption}
\usepackage{cite}
\usepackage{graphicx}
\usepackage{amsmath}
\usepackage{amssymb}
\usepackage{color}
\usepackage{breqn}
\usepackage{lipsum,mwe}
\usepackage{multirow}
\usepackage{array}
\usepackage{mathrsfs}
\usepackage{enumerate}

\begin{document}

\begin{center}
{\Large\bf An Alternative Approach to the Exact Solution of FRW-Type Spacetime with a Generalized Chaplygin Gas}
\\[15 mm]
D. Panigrahi\footnote{ Netaji Nagar Day College, Kolkata 700092, India \textit{and also} Relativity and Cosmology Research Centre, Jadavpur University, Kolkata 700032,
 e-mail:
dibyendupanigrahi@yahoo.co.in; dpanigrahi@nndc.ac.in  }\\[10mm]

\end{center}

\begin{abstract}
 The generalized Chaplygin gas model, characterized by the equation of state
 $p = - \frac{B}{\rho^{\alpha}}$,  is investigated within the framework of a Robertson-Walker spacetime. The resulting field equations governing this model are highly non-linear in the scale factor, forming the central focus of this work. Previous studies have employed this equation to describe both a dust-dominated universe and an accelerating universe in two extreme cases. However, the time evolution of the scale factor between these two extremal cases remains unclear. To address this short coming, we have employed a first-order approximation of the key equation and subsequently derived exact time-dependent solutions for the scale factor. The obtained solution converges to the $\Lambda$CDM model at large scale factors and exhibits the desirable feature of an acceleration flip. A detailed analysis of the flip time has been conducted, providing both analytical and graphical insights. The parameters of the model have been constrained using the Hubble-$57$ dataset. The present age of the universe has also been calculated.  A comparison of the results from both the theoretical and observational approaches reveals remarkable consistency, with the theoretical graph of $H(z)$ vs. $z$ closely aligning with the best-fit graph obtained from the Hubble-$57$ dataset. Furthermore, the entire scenario has been examined within the context of the well-known Raychaudhuri equation, offering a broader perspective and comparison with previous results.
\end{abstract}


\section{ Introduction}

Three discoveries in the last century have radically transformed our understanding of the universe. First, contrary to Einstein's concept of a static universe, Hubble and Slipher (1927) demonstrated that the universe is expanding. Second, the discovery of the cosmic microwave background radiation (CMBR) and the analysis of primordial nucleosynthesis in the 1960s revealed that the universe began in a hot, dense state and has been expanding for the last $13.5$ billion years. Third, assuming the validity of Einstein's theory and the Friedmann-Robertson-Walker (FRW) cosmological model, observations of type Ia supernovae as standardized candles suggest~\cite{res} that the universe is currently undergoing accelerated expansion, with baryonic matter contributing only about $5$\% of the total energy budget. Later data from CMBR probes~\cite{spe} further support this conclusion.

This evidence has motivated a significant portion of the cosmology community to seek explanations for this acceleration of the universe. The central question in this field is identifying the mechanisms responsible for triggering this late-time inflation. Researchers are broadly divided into two camps: those advocating modifications to Einstein's original theory, and those introducing exotic forms of matter such as a cosmological constant or a quintessential scalar field.

The popular explanation involving a cosmological constant faces severe theoretical challenges. The absence of acceleration at redshifts $ z \geq 1 $ requires the cosmological constant to be approximately $120 $ orders of magnitude smaller than its natural value based on the Planck scale~\cite{cop}. On the other hand, the alternative hypothesis of a quintessential field~\cite{sam} lacks a solid theoretical foundation, as no existing theory can predict or explain the existence of such a scalar field without violating realistic energy conditions. Moreover, it is not possible to derive such a scalar field from first principles of physics.

Other alternatives include k-essence~\cite{sch}, tachyons~\cite{gib}, phantom fields~\cite{eli}, and quintom models~\cite{guo}. These challenges have spurred renewed interest among relativists, field theorists, astrophysicists, and astroparticle physicists, both theoretical and experimental, in addressing the issues posed by recent extragalactic observations. The aim is to seek explanations that do not rely on ad hoc exotic scalar fields but are instead based on sound physical principles. Proposed alternatives include higher-curvature theories, axionic fields, and Brans-Dicke fields. Some researchers have approached the problem from a purely geometric perspective, in line with Einstein's principles. For example, Wanas~\cite{wan} introduced torsion, while Neupane~\cite{neu} modified the spacetime geometry with a warped factor in $5D$ brane cosmology. Additionally, the inclusion of extra spatial dimensions, as predicted by string theory~\cite{dp1,dp2,sah}, has been explored.

While torsion-inspired inflation models have certain desirable features, the main issue with Wanas' model is that the geometry is no longer Riemannian. Furthermore, several researchers~\cite{kra} have questioned the homogeneity assumption itself, arguing that accelerating models and exotic matter fields are only necessary within the context of FRW cosmology.

Among the various alternatives proposed to explain the observed acceleration of the universe, one model that has attracted significant attention is the introduction of a Chaplygin-type gas as a new matter field to mimic dark energy. This matter field was later generalized by introducing an arbitrary constant as an exponent to the mass density~\cite{bent}, and it is now commonly referred to as the generalized Chaplygin gas (GCG)~\cite{bento, bert, wu, thak}. The GCG cosmological model describes the dynamics of both dark matter and dark energy. Its equation of state (EoS) can also be derived from the Nambu-Goto action for d-branes moving in a $(d+2)$-dimensional spacetime in the light-cone parameterization~\cite{bord}. Furthermore, the GCG is notable as it is the only fluid model, to date, that admits a supersymmetric generalization~\cite{hop,jac}. From a cosmological perspective, the GCG provides a potential unification of dark matter and dark energy, offering a possible solution to the so-called cosmic coincidence problem~\cite{zla}. This has motivated further studies of the GCG model, which dynamically unifies dark matter and dark energy. Such models are often referred to as unified dark matter (UDM) models.

In this work, we revisit the dynamics of the FRW model by considering the GCG as the matter field. We explore some previously unexplored aspects of the model and derive interesting results. In Section-2, we present the field equations and the equation of state. Section-3 provides the mathematical formulation, leading to a hypergeometric solution. The deceleration parameter, an effective equation of state, redshift at flip time, and jerk parameter are also derived here. Our model asymptotically approaches $\Lambda$CDM
 at future cosmic times. The evolution is also illustrated graphically. Additionally, the parameters are constrained using the Hubble $57$ data points. In Section-4, we highlight an interesting feature of our analysis, where the first-order approximation of our key equation has been considered. Using this approach, we derive exact solutions of scale factor.  A detailed analysis of the flip time is also carried out, both analytically and graphically. Section-5 examines the conclusions within the framework of the well-known Raychaudhuri equation. The paper concludes with a discussion in Section-6.

\section{ Field Equations}

We consider a spherically symmetric flat homogeneous spacetime given by
\begin{equation}\label{eq:1}
  ds^{2}= dt^{2}- a^2(t)~\{dr^{2}+r^{2}\left( d\theta^{2}+\sin^{2}\theta d\phi^{2} \right) \}
\end{equation}

where the scale factor, $a(t)$  depends on time only.

 \vspace{0.1 cm}
A comoving coordinate system is taken
such that $ u^{0}=1, u^{i}= 0 ~(i = 1, 2,3)$ and $g^{\mu
\nu}u_{\mu}u_{\nu}= 1$ where $u_{i}$ is the 4-velocity. The
energy momentum tensor for a dust distribution in the above
defined coordinates is given by

\begin{equation} \label{eq:2}
T^{\mu}_{\nu} = (\rho + p)\delta_{0}^{\mu}\delta_{\nu}^{0} -
p\delta_{\nu}^{\mu}
\end{equation}
where $\rho(t)$ is the matter density  and $p(t)$ the isotropic
 pressure.
The independent field equations for the metric \eqref{eq:1} and the energy
momentum tensor \eqref{eq:2} are given by

\begin{equation}\label{eq:3}
3\frac{\dot{a}^2}{a^2} = 3H^2=\rho
\end{equation}
\begin{equation}\label{eq:4}
 2 \frac{\ddot{a}}{a} + \frac{\dot{a}^2}{a^2} = 2(\dot{H} + H) = - p
\end{equation}

 \vspace{0.2 cm}

From the the  conservation law we get

\begin{equation}\label{eq:5}
\nabla_{\nu}T^{\mu \nu}= 0
\end{equation}

which, in turn, yields for the line element \eqref{eq:1}

\begin{equation}\label{eq:6}
\dot{\rho} + 3 H (\rho + p) = 0
\end{equation}

At this stage we consider a generalised
Chaplygin type of gas (GCG) obeying an equation of state ~\cite{bento}

\begin{equation}\label{eq:7}
p =  -\frac{B}{\rho^\alpha}
\end{equation}
where the constant $ B > 0 $ is related to the dimensionless parameter $ B_s $ as $ B_s = \frac{B^{\frac{1}{1+\alpha}}}{\rho_0}$, where $ B_s > 0 $, and $ \rho_0 $ represents the present density of the GCG. The parameter $\alpha$ is a positive constant within the range $ 0 < \alpha \leq 1 $. To ensure that the square of the sound speed, $c_s^2 = \alpha \frac{B}{\rho^{\alpha+1}}$, does not exceed the square of the speed of light $(c^2)$ gives the condition $ 0 < \alpha < \frac{c^2 \rho^{\alpha+1}}{B}$.
As the universe evolves and the density $\rho$ decreases, the right-hand side of this inequality also decreases, requiring $\alpha$ to take smaller values. This implies that in the late stages of the universe, smaller values of $\alpha$ are expected. This behavior aligns with the GCG model, where the equation of state softens at low densities, reflecting the universe's transition to a dark energy-dominated phase. Additionally, for $ B > 0 $, the sound speed can also be expressed as $ c_s^2 = -\alpha \frac{p}{\rho}$. To ensure the stability of perturbations, the sound speed squared must remain subluminal $( c_s^2 \leq 1 )$ and physically meaningful. If $\alpha \geq 1$, the sound speed can become superluminal $(c_s^2 > 1)$ in certain cases, violating causality. Therefore, the constraint $\alpha < 1 $ is necessary to maintain physical viability (causality and stability), observational consistency, and a smooth evolution of the universe in the GCG framework.

 Now with the help of eqs~\eqref{eq:6} \& \eqref{eq:7} a little mathematics shows that
the expression for density comes out to be

\begin{equation}\label{eq:8}
\rho (a) =
a^{-3}\left[3(1+\alpha)\int B a^{3(1+\alpha)
- 1}~ da + c \right]^{\frac{1}{1+\alpha}}
\end{equation}

where $c$ is an integration constant. The above eq.~\eqref{eq:8}is written in terms of $z$ as
\begin{equation}\label{eq:8a}
\rho (z) = (1+z)^{3}\left[-3B(1+\alpha)\int  \frac{dz}{(1+z)^{3(1+\alpha)
+1}} + c \right]^{\frac{1}{1+\alpha}}
\end{equation}
which yields a first integral as
\begin{equation}\label{eq:9}
\rho = \left[ B +
c z^{3(1+\alpha)}\right]^{\frac{1}{1+\alpha }}
\end{equation}

Plugging in the expression of $\rho$ from eqs~\eqref{eq:3} and \eqref{eq:9} we
finally get

\begin{equation}\label{eq:10}
3 \frac{\dot{a}^{2}}{a^{2}} = \ \left[ B +
c z^{3(1+\alpha)}\right]^{\frac{1}{1+\alpha }}
\end{equation}

\section{Cosmological dynamics}

It is quite difficult to obtain the exact temporal behavior of the scale factor, $a(t)$, from eq.~\eqref{eq:10} in a closed form because the integration yields elliptical solutions, resulting in hypergeometric series. However, eq.~\eqref{eq:10} still provides significant information under extremal conditions, as briefly discussed below.

\vspace{0.5 cm}

\subsection{\textbf{Deceleration Parameter:}}

At the early stage of cosmological evolution, when the scale factor $ a(t)$ is relatively small, the second term on the right-hand side of eq.~\eqref{eq:10} dominates. This behavior has been extensively discussed in the literature~\cite{bento}, and therefore, we will only briefly address it here. From the expression of the deceleration parameter $q $, we obtain,

\begin{equation}\label{eq:11}
q = -\frac{1}{H^{2}}\frac{\ddot{a}}{a}= \frac{d}{dt} \left(H^{-1}
\right) - 1 = \frac{1}{2} + \frac{3}{2}\frac{p}{\rho}
\end{equation}

where $H$ is the Hubble constant. With the help of the Equation of State (EoS) given
by \eqref{eq:7} we find

\begin{equation}\label{eq:12}
q = \frac{1}{2}- \frac{3B}{2}\frac{1}{\rho^{\alpha +1}}
\end{equation}

which in terms of scale factor via eq.~\eqref{eq:9} gives

\begin{equation}\label{eq:13}
q = \frac{1}{2}- \frac{3B}{2}\left[B +
\frac{c}{ a^{3(1+\alpha)}}\right]^{-1}
\end{equation}

Again at flip time, i.e. when $q = 0$ the scale factor becomes

\begin{equation}\label{eq:13a}
a = \left(\frac{c}{2B}\right)^{\frac{1}{3(1+\alpha)}}
\end{equation}
shows that for small $\alpha$, the scale factor $a$ becomes large as expected. On the other hand, to achieve acceleration ($q < 0$), it follows that $a > \left(\frac{c}{2B}\right)^{\frac{1}{3(1+\alpha)}}$.
Now, in terms of the redshift, defined by $1+z = \frac{1}{a}$, we can express equation \eqref{eq:12} in the following form

\begin{equation}\label{eq:13b}
q = \frac{1}{2}- \frac{3B}{2}
\left[B + c (1+z)^{3(1+\alpha)}\right]^{-1}
\end{equation}
 To constrain the parameters, let us consider $\Omega_m = \frac{c}{\rho_0^{1+\alpha}}$. Now using eq.~\eqref{eq:13b}, we get

\begin{equation}\label{eq:15}
q = \frac{1}{2} -\frac{3}{2}\frac{1-\Omega_m}{\Omega_m}\left[\frac{1-\Omega_m}{\Omega_m} + (1+z)^{3(1+\alpha)}\right]^{-1}
\end{equation}

\begin{figure}[ht]
\begin{center}
  \includegraphics[width=10cm]{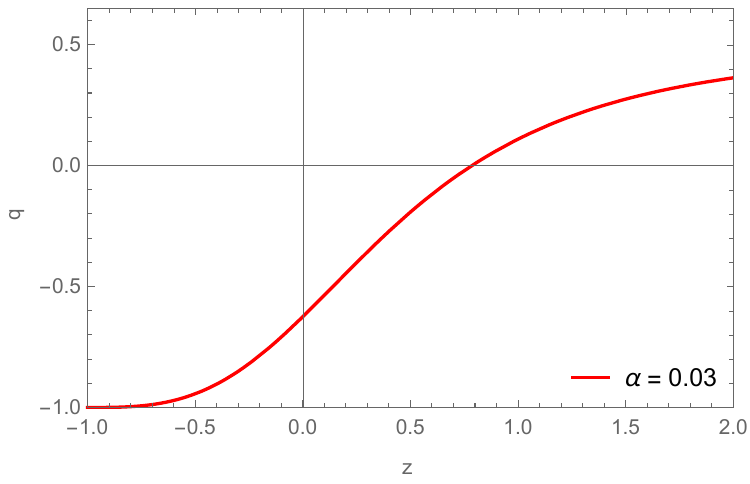}
  \caption{
  \small\emph{The variation of $q$ with $z$ is shown in this figure.    }\label{qz}
    }
\end{center}
\end{figure}
Again at flip time, i.e., when $q = 0$ the redshift parameter  $z_{f}$ from eq.~\eqref{eq:13b}

\begin{equation}\label{eq:16a}
z_{f} =\left(\frac{2B}{c} \right)^{\frac{1}{3(1+\alpha)} }-1
\end{equation}
and from eq.~\eqref{eq:15}, we get
\begin{equation}\label{eq:16b}
z_{f} = \left[\frac{2(1-\Omega_m)}{\Omega_m} \right]^{\frac{1}{3(1+\alpha)}} - 1
\end{equation}
where $z_{f}$ signifies the sign change of the deceleration parameter. For the universe to be accelerating at the present epoch (\textit{i.e.}, at $z = 0$), we require $z_{f} > 0$. From eq.~\eqref{eq:16a}, this condition implies that $ B > \frac{c}{2}$, or equivalently, $\frac{B_s^{1+\alpha}}{\Omega_m} > \frac{1}{2}$, which provides a critical constraint for the model.
Additionally, from eq.~\eqref{eq:16b}, we deduce $\Omega_m < \frac{2}{3}$, which aligns with current observational constraints on $\Omega_m$ (in our case, $\Omega_m = 0.2443$). These results demonstrate consistency with an accelerating universe, reinforcing the validity of the model in light of current observations.

\hspace{-0.6cm}As the universe expands the energy density  $\rho$ decreases with time such that the
last term in the eq.~\eqref{eq:12} increases indicating  a sign flip when the density attains a critical value given by
\begin{equation}\label{eq:18a}
 \rho_{f} = (3B )^{\frac{1}{1+ \alpha}}
\end{equation}
This flip density $\rho_{f}$ depends on the exponent $\alpha$ and  parameter $B$.

\hspace{-0.6cm}Now we discuss the extremal cases to understand the evolution of the universe.

 \begin{itemize}
    \item[(i)]  In the early phase, when the redshift $z$ is very high, eq.~\eqref{eq:15} reduces to:
    \begin{equation}\label{eq:19}
        q = \frac{1}{2}
    \end{equation}
    This represents a dust-dominated universe for the generalized Chaplygin gas. The positive value of $q$ indicates a deceleration phase. Fig.-\ref{qz} shows that at high $z$, $q $ approaches $0.5$.

    \item[(ii)] At present, when the redshift $z = 0$, eq.~\eqref{eq:15} reduces to:
    \begin{equation}\label{eq:19a}
        q = -1 + \frac{3 \Omega_m}{2}
    \end{equation}
    This implies $0 > q > -1$, where $q$ is negative, representing an accelerating universe during this phase, as shown in fig.-\ref{qz}. In this case, later we get $q \approx -0.625$, which is less than zero but greater than $-1$, indicating acceleration.

    \item[(iii)] In the later epoch of evolution, \textit{i.e.}, when the universe has reached a large size, eq.~\eqref{eq:15} gives:

 \begin{equation}\label{eq:20}
        q = -1
    \end{equation}
    This represents a pure $\Lambda$CDM model which shows also in fig-\ref{qz}.
 \end{itemize}

It is to be mentioned that a central challenge in cosmology is the cosmic coincidence problem — the question of why the energy densities of dark matter and dark energy are of the same order precisely in the current epoch. While models like quintessence and k-essence attempt to address this through evolving scalar fields with canonical or non-canonical kinetic terms, they typically treat dark matter and dark energy as independent components. Although these frameworks offer tracker or attractor solutions to reduce fine-tuning, they leave the coincidence problem only partially resolved.

In contrast, the Generalized Chaplygin Gas (GCG) model naturally addresses this issue by unifying dark matter and dark energy as different dynamical regimes of a single cosmic fluid. Its equation of state evolves smoothly from a pressureless, matter-like phase at early times to a negative-pressure, dark energy-like state at late times, inherently explaining the observed balance of densities without requiring separately tuned components.

However, despite this conceptual advantage, the GCG model faces notable limitations — particularly, tensions with structure formation data in its original formulation and the need for stringent parameter constraints to remain consistent with cosmic microwave background anisotropies and large-scale structure observations. Nevertheless, it continues to offer a distinctive and physically motivated alternative within the broader landscape of cosmological models aiming to explain the universe’s accelerated expansion.

\subsection{Effective Equation of State:}

\begin{figure}[ht]
\begin{center}
  \includegraphics[width=10cm]{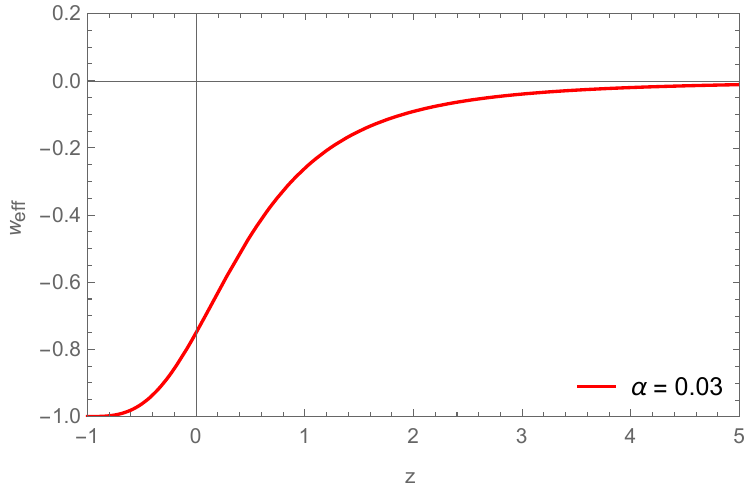}
  \caption{
  \small\emph{The variation of $w_e$ with $z$ is shown in this figure.    }\label{wz}
    }
\end{center}
\end{figure}

It further gives the \emph{effective} EoS using the eq. \eqref{eq:10},
\begin{equation}\label{eq:21}
w_{\text{eff}}  =  \frac{p}{\rho} =  - \frac{1-\Omega_m}{ 1 - \Omega_m +\Omega_m(1+z)^{3(1+\alpha)}}
\end{equation}
\begin{itemize}
    \item[(i)]~~ When the  redshift $z$  is very high enough representing the early phase of the universe, the  eq.~\eqref{eq:21} reduces to $w_{\text{eff}} = 0$ implying a dust dominated universe as shown in  fig.-\ref{wz} also. \\
 \item[(ii)]~~At present the redshift $z = 0$ and the eq.~\eqref{eq:21} reduces to
\begin{equation}\label{eq:19a}
w_{\text{eff}} = -1 + \Omega_m
\end{equation}
which represents accelerating universe at present epoch, later we get  $w_{\text{eff}} \approx -0.75$.
\item[(iii)]~~  At the late stage of evolution, {\it i.e.}, for a large size of the universe,  the effective EoS is obtained  from the eq.~\eqref{eq:21} as
\begin{equation}\label{eq:20}
w_{\text{eff}} = -1
\end{equation}
representing a $\Lambda$CDM model.
\end{itemize}

The evolution of $w_{eff}$ indicates that the universe transitions from a dust-dominated phase to a $\Lambda$CDM phase, as shown in fig.-\ref{wz}.

\subsection{Jerk parameter:}
\vspace{0.3 cm}

The jerk parameter, a dimensionless third derivative of the scale factor $a(t)$ with respect to cosmic time $t$, provides a simple approach to identify departures from the concordance $\Lambda$CDM model. It is defined as~\cite{blan,rap}

\begin{equation}\label{eq:18b}
j = \frac{dq}{dt} = - \frac{1}{a H^3}\frac{d^3a}{dt^3}
\end{equation}

Now, the jerk parameter $j$ can be written  in terms of deceleration parameter $q $ as

\begin{equation}\label{eq:18c}
 j(q) =   q(2q + 1) + (1 + z)\frac{dq}{dz}
\end{equation}

Blandford \textit{et al.}~\cite{blan} described how the jerk parameterization provides an alternative and convenient method to describe cosmological models close to the concordance $\Lambda$CDM model. A notable feature of the jerk parameter, $j$, is that for the $\Lambda$CDM model, $j = 1$ (constant) always holds.
It is important to note that Sahni \textit{et al.}~\cite{sah1, alam} emphasized the significance of $j$ in distinguishing between different dark energy models. Any deviation from $j = 1$, much like deviations from the effective equation of state parameter $w_{\text{eff}} = -1$ in more conventional dynamical approaches, would indicate a departure from the $\Lambda$CDM model. The simplicity of the jerk formalism thus provides an effective way to constrain deviations from the $\Lambda$CDM model.

Now using eqs~\eqref{eq:15} and \eqref{eq:18c} we get

\begin{dmath}\label{eq:18d}
 j(q) = \left\{ \frac{1}{2} - \frac{\frac{3}{2} \left( \frac{1 - \Omega_m}{\Omega_m} \right)}{ \frac{1 - \Omega_m}{\Omega_m} +(1+z)^{3(1+\alpha)}}  \right\} 
 \left\{ 2 - \frac{3 \left( \frac{1 - \Omega_m}{\Omega_m} \right)}{ \frac{1 - \Omega_m}{\Omega_m} +(1+z)^{3(1+\alpha)}} \right\}   \\+
\frac{\frac{9}{2} (1+\alpha)\left( \frac{1 - \Omega_m}{\Omega_m} \right)}{ \left\{ \frac{1 - \Omega_m}{\Omega_m} +(1+z)^{3(1+\alpha)}\right\}^{2}}  (1+z)^{3(1+\alpha)}
\end{dmath}

\begin{figure}[ht]
\begin{center}
  \includegraphics[width=10cm]{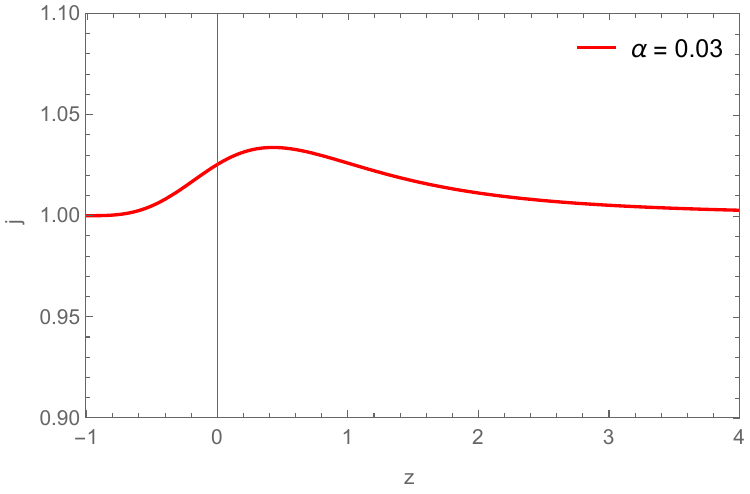}
  \caption{
  \small\emph{The variation of $j$ with $z$ is shown in this figure.    }\label{jz1}
    }
\end{center}
\end{figure}

The eq.~\eqref{eq:18d} is quite involved, so we now examine its behavior in three extremal cases:

\begin{itemize}
\item[(i)] In the early phase, when the redshift $z$ is very high, eq.~\eqref{eq:18d} simplifies to $j = 1$.

This corresponds to the matter-dominated era, where the dynamics are governed by the standard cosmological model.
\item[(ii)] At the present epoch, when the redshift $z = 0$, eq.~\eqref{eq:18d} reduces to
$ j = \frac{1}{2} \left\{ 2 + 9 \alpha \Omega_m (1 - \Omega_m) \right\}$.

For $\alpha = 0.03$ and $\Omega_m = 0.2443$, this yields $j \approx 1.025 $,
which is slightly higher than unity. This small deviation reflects the influence of the GCG’s dark energy-like behavior beginning to emerge in the current epoch, as the universe transits from matter domination to accelerated expansion.
\item[(iii)] In the far future, \textit{i.e.}, when the universe has expanded to a large size, eq.~\eqref{eq:18d} again approaches $ j = 1$.
This corresponds to a de Sitter phase dominated by a cosmological constant.
\end{itemize}
Since the jerk parameter $j$ remains nearly unity throughout the entire cosmic evolution, it indicates that the universe's expansion in the GCG model effectively mimics the behavior of a $\Lambda$CDM cosmology — exhibiting matter domination at early times and approaching a de Sitter phase at late times — without the need for an explicit cosmological constant term. The graphical representation of the jerk parameter $j$ as a function of cosmic time $t$, shown in fig.~\ref{jz1}, further corroborates this behavior. This result remains consistent with current observational constraints on the expansion history of the universe.

\vspace{0.5 cm}

\subsection{Observational Constraints on  the  Model Parameters:}
\vspace{0.3 cm}

In this section, the Hubble-$57$ data~\cite{sha} will be utilized to analyze the cosmological model by estimating the constraints imposed on the model parameters. The Hubble parameter, $H(z)$, at a given redshift $z$, can be measured using two methods:
\begin{enumerate}[(i)]
    \item Estimations of $H(z)$ from differential ages (DA) $\Delta t$ of galaxies~\cite{zhang}-\cite{mor3}.
    \item Extraction of $H(z)$ from line-of-sight BAO data~\cite{gaz}-\cite{fon}, including analyses of the correlation functions of luminous red galaxies~\cite{oka, chu}.
\end{enumerate}

\begin{table}[h!]
\small
  \centering
  \caption{The latest Hubble parameter measurements $H(z)$ (in units of $km s^{-1}Mpc^{-1}$ ) and their errors $\sigma$  at redshift $z$ obtained from the differential age method (CC). }\vspace{0.5 cm}
  \label{tab:table1}
\begin{tabular}{|c|c|c|c||c|c|c|c|}  \hline
   \multicolumn{4}{|c||}{DA Method} & \multicolumn{4}{c|}{BAO Method} \\ \hline
  z & H(z) & $\sigma$ &  Reference & z & H(z) & $\sigma$ &  Reference \\   \hline \hline
  0.0700 & 69.00 & $\mp 19.6 $& \ ~\cite{zhang} &  0.24 & 79.69 & $\mp 2.99 $& \ ~\cite{gaz}\\
  0.0900 & 69.60 & $\mp 12.00 $ &   ~\cite{ster} &0.30 & 81.70 & $\mp 6.22 $ &   ~\cite{oka}\\
  0.1200 & 68.60 & $\mp 26.20$ &  ~\cite{zhang} & 0.31 & 78.18 & $\mp 4.74$ &  ~\cite{wan}\\
  0.1700 & 83.00 & $\mp 8.00 $&   ~\cite{ster} & 0.34& 83.80 & $\mp 3.66 $&   ~\cite{gaz}\\
  0.1791 & 75.00 & $ \mp 4.00$ &  ~\cite{mor1} & 0.35 & 82.70 & $ \mp 9.10$ &  ~\cite{chu}\\
  0.1993 & 75.00 & $\mp 5.00$ &  ~\cite{mor1} & 0.36 & 79.94 & $\mp 3.38$ &  ~\cite{wan}\\
  0.2000 & 72.90 & $\mp 29.60 $&  ~\cite{zhang} & 0.38 & 81.50 & $\mp 1.90 $&  ~\cite{ala} \\
  0.2700 & 77.00 &$ \mp 14.00$ &  ~\cite{ster} & 0.40 & 82.04 & $\mp 2.03 $& ~\cite{wan}\\
  0.2800 & 88.80 & $\mp 36.60 $& ~\cite{zhang} & 0.43 & 86.45 & $\mp 3.97 $& ~\cite{gaz}\\
  0.3519 & 83.00 & $\mp 14.00$ &  ~\cite{mor1} & 0.44 & 82.60 & $\mp 7.80$ &  ~\cite{bla}\\
  0.3802 & 83.00 & $\mp 13.50 $&  ~\cite{mor2} & 0.44 & 84.81 & $\mp 1.83$&  ~\cite{wan}\\
  0.4000 & 95.00 &$ \mp 17.00$ &   ~\cite{ster}& 0.48 & 87.79 &$ \mp 2.03$ &   ~\cite{wan}\\
  0.4004 & 77.00 &$ \mp 10.20 $&   ~\cite{mor2} & 0.51 & 90.40 &$ \mp 1.90 $&   ~\cite{ala} \\
  0.4247 & 87.10 & $\mp 11.20 $&   ~\cite{mor2} & 0.52 & 94.35 & $\mp 2.64 $&   ~\cite{wan}\\
  0.4497 & 92.80 &$ \mp 12.90 $&  ~\cite{mor2} & 0.56 & 93.34 &$ \mp 2.30 $&  ~\cite{wan} \\
  0.4700 & 89.00 &$ \mp 34.00 $&  ~\cite{rats} & 0.57 & 87.60 &$ \mp 7.80 $&  ~\cite{chu}\\
  0.4783 & 80.90 &$\mp 9.00 $&   ~\cite{mor2} & 0.57 & 96.80 &$\mp 3.40 $&   ~\cite{del}\\
  0.4800 & 97.00 & $\mp 62.00$ &  ~\cite{ster} & 0.59 & 98.48 & $\mp 3.18$ &  ~\cite{wan}\\
  0.5929 & 104.00 &$ \mp 13.00 $&  ~\cite{mor1} &  0.60 & 87.90 & $\mp 6.10$ &  ~\cite{bla}\\
  0.6797 & 92.00 & $\mp 8.00$ &  ~\cite{mor1} & 0.61 & 97.30 & $\mp 2.10$ &  ~\cite{ala}\\
  0.7812 & 105.00 & $\mp 12.00$ &  ~\cite{mor1} & 0.64 & 98.82 & $\mp 2.98$ &  ~\cite{wan}\\
  0.8754 & 125.00 & $\mp 17.00$ & ~\cite{mor1} & 0.73 & 97.30 & $\mp 7.00$ & ~\cite{bla} \\
  0.8800 & 90.00 & $\mp 40.00 $ & ~\cite{ster}& 2.30 & 224.00 & $\mp 8.60 $ & ~\cite{bus} \\
  0.9000 & 117.00 &$ \mp 23.00 $&   ~\cite{ster} & 2.33 & 224.00 &$ \mp 8.00 $&   ~\cite{bau}\\
  1.0370 & 154.00 &$\mp 20.00 $&   ~\cite{mor1} & 2.34 & 222.00 &$\mp 8.50 $&   ~\cite{del}\\
  1.3000 & 168.00 & $\mp 17.00$ &  ~\cite{ster} & 2.36 & 226.00 & $\mp 9.30$ &  ~\cite{fon}\\ \cline{5-8}
  1.3630 & 160.00 &$ \mp 33.60 $&  ~\cite{mor3}\\
  1.4300 & 177.00 & $\mp 18.00$ &   ~\cite{ster}\\
  1.5300 & 140.00 & $\mp 14.00$ & ~\cite{ster}\\
  1.7500 & 202.00 & $\mp 40.00$ &  ~\cite{ster} \\
  1.9650 & 186.50 & $\mp 50.40 $ &  ~\cite{mor3} \\
   \cline{1-4}
\end{tabular}
\end{table}
The Hubble parameter depending on the differential ages as a function of  redshift $ z$ can be written in the form of

\begin{equation}\label{eq:20a}
H(z) = -\frac{1}{1+z} \frac{dz}{dt} \simeq  -\frac{1}{1+z} \frac{\vartriangle z}{\vartriangle t}
\end{equation}
Therefore, $H(z)$ can be determined directly from eq.~\eqref{eq:20a} once $\frac{dz}{dt}$ is known~\cite{sei}. Using the present value of the scale factor normalized to unity, \emph{i.e.}, $a = a_0 = 1$, we obtain a relation between the Hubble parameter and the redshift parameter $z$.
If $\rho_0$ denotes the density at the present epoch, then the well-known density parameter is written as $\Omega_m = \frac{c}{\rho_0^{1+\alpha}}$~\cite{seth}. Now, using eq.~\eqref{eq:10}, we can express the three-dimensional spatial matter density as

\begin{equation}\label{eq:21a}
\rho = \rho_0 \left \{1-\Omega_m + \Omega_m \left(1+z \right)^{3(1+\alpha)} \right \}^{\frac{1}{1+\alpha}}
\end{equation}

\begin{figure}[ht]
\begin{center}
  \includegraphics[width=10cm]{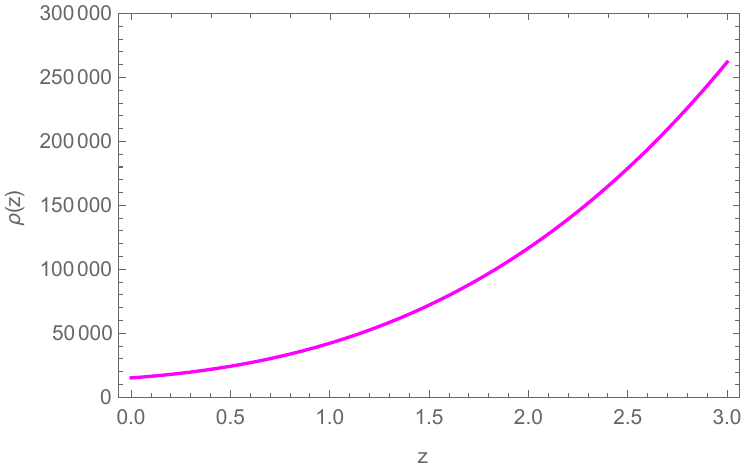}
  \caption{
  \small\emph{The variation of $\rho$ with $z$ is shown in this figure.    }\label{rz}
    }
\end{center}
\end{figure}
\begin{figure}[ht]
    \centering
    \subfigure[\small\emph{Best fit graph using Hubble $57$ data}]{
        \includegraphics[width=5.5cm]{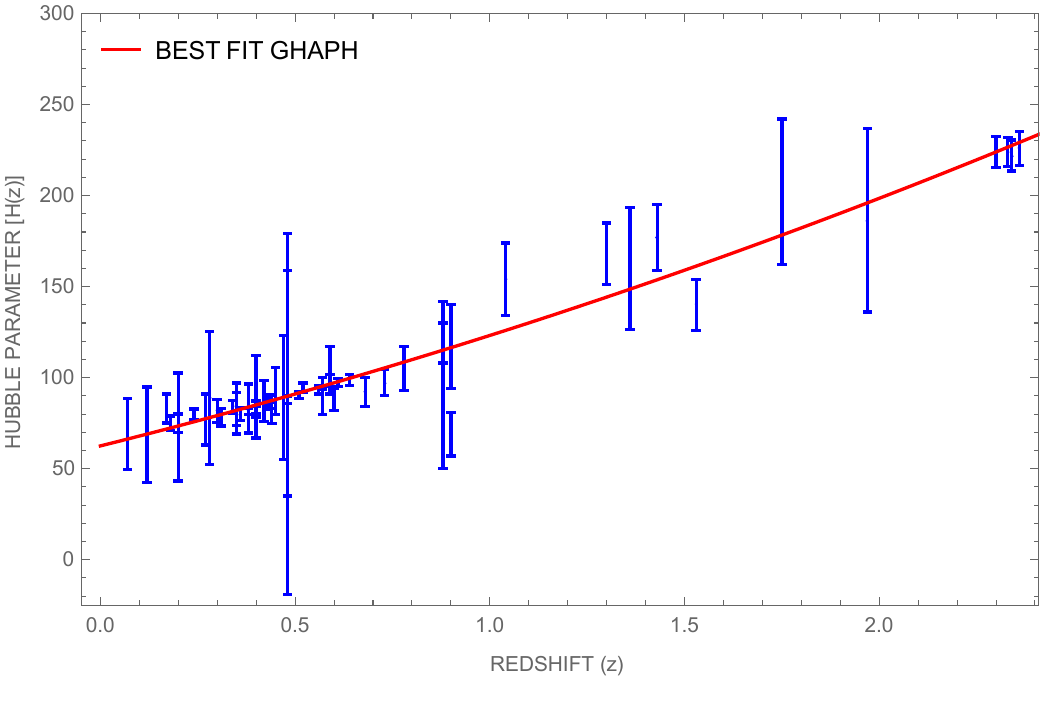}
        \label{hz1}
    }
    ~~~
    \subfigure[\small\emph{Best fit graph with Theoretical graph}]{
        \includegraphics[width=5.5cm]{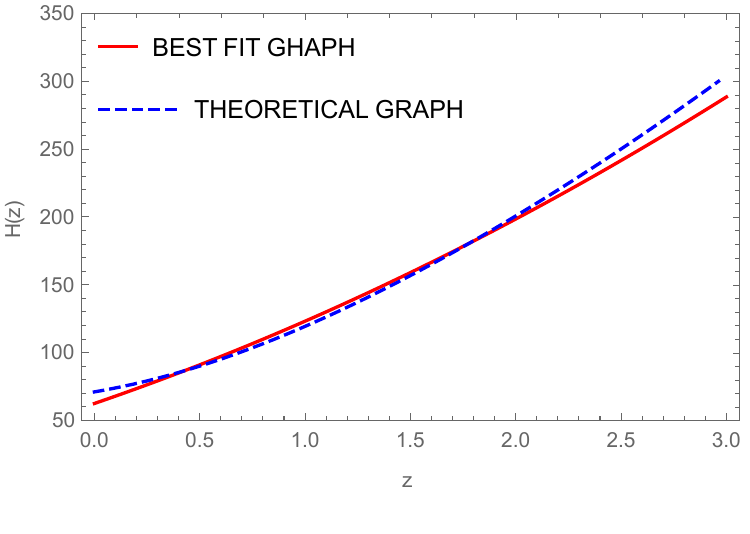}
        \label{hz2}
    }
    \caption[\small Optional caption for list of figures]{
        \small\emph{$H(z)$ vs $z$}
    }
    \label{fig:hz}
\end{figure}

The fig.-\ref{rz} illustrates the evolution of the universe's density as a function of redshift, $z$. It shows that the matter density decreases as $z$ decreases, consistent with expectations from cosmological models where matter dilutes as the universe expands. Now the Hubble parameter
\begin{equation}\label{eq:21b}
 H(z) = H_0 \left \{1-\Omega_m + \Omega_m \left(1+z \right)^{3(1+\alpha)} \right \}^{\frac{1}{2(1+\alpha)}}
\end{equation}
where $H_0 = \left(\frac{\rho_0}{3} \right)^{\frac{1}{2}}$ represents the present value of the Hubble parameter. The equation~\eqref{eq:21b} describes the evolution of the Hubble parameter $H(z)$ as a function of the redshift parameter $z$. In fig.-\ref{hz1}, we present a best-fit curve of the redshift $z$ against the Hubble parameter $H(z)$ using the Hubble $57$ data points. Furthermore, in fig.-\ref{hz2}, we compare the best-fit graph with the graph obtained from eq.~\eqref{eq:21b}. These two graphs nearly coincide throughout the evolution, indicating that the behavior of our model is in good agreement with the observational data.

The apparently small uncertainty of the measurement naturally increases its weightage in
estimating $\chi^2$ statistics. We define here the   $\chi^2$ as
\begin{equation}\label{eq:20b}
\chi_{H}^2 = \sum_{i=1}^{30} \frac{[H^{obs}(z_i) - H^{th} (z_i, H_0, \theta]^2}{\sigma^2_H(z_i)}
\end{equation}
Here, $H^{\text{obs}}$ represents the observed Hubble parameter at $z_i$, and $H^{\text{th}}$ is the corresponding theoretical Hubble parameter given by eq.~\eqref{eq:21b}. Additionally, $\sigma_H(z_i)$ denotes the uncertainty for the $i$-th data point in the sample, and $\theta$ represents the model parameter. In this work, we utilize the observational $H(z)$ dataset, which consists of $57$ data points spanning the redshift range $0.07 \leq z \leq 2.36$, extending beyond the redshift range covered by type Ia supernova observations. It is important to note that the confidence levels $1\sigma$ ($68.3\%$), $2\sigma$ ($95.4\%$), and $3\sigma$ ($99.7\%$) correspond to $\Delta \chi^2$ values of $2.3$, $6.17$, and $11.8$, respectively, where $ \Delta \chi^2 = \chi^2(\theta) - \chi^2(\theta^*) $ and $\chi^2_m$ is the minimum value of $\chi^2$. An important quantity used in the data fitting process is
\begin{equation}\label{eq:20c}
\overline{\chi^2} =  \frac{\chi_m^2}{dof}
\end{equation}
where, the subscript {\it dof} represents the degrees of freedom, defined as the difference between the total number of observational data points and the number of free parameters. If $\frac{\chi_m^2}{\textit{dof}} \leq 1$, it indicates a good fit, implying that the observed data are consistent with the considered model.

\begin{figure}[h!]
    \centering 
    \begin{tabular}{l l}
        \parbox{2in}{%
            \includegraphics[width=2in,height=2in]{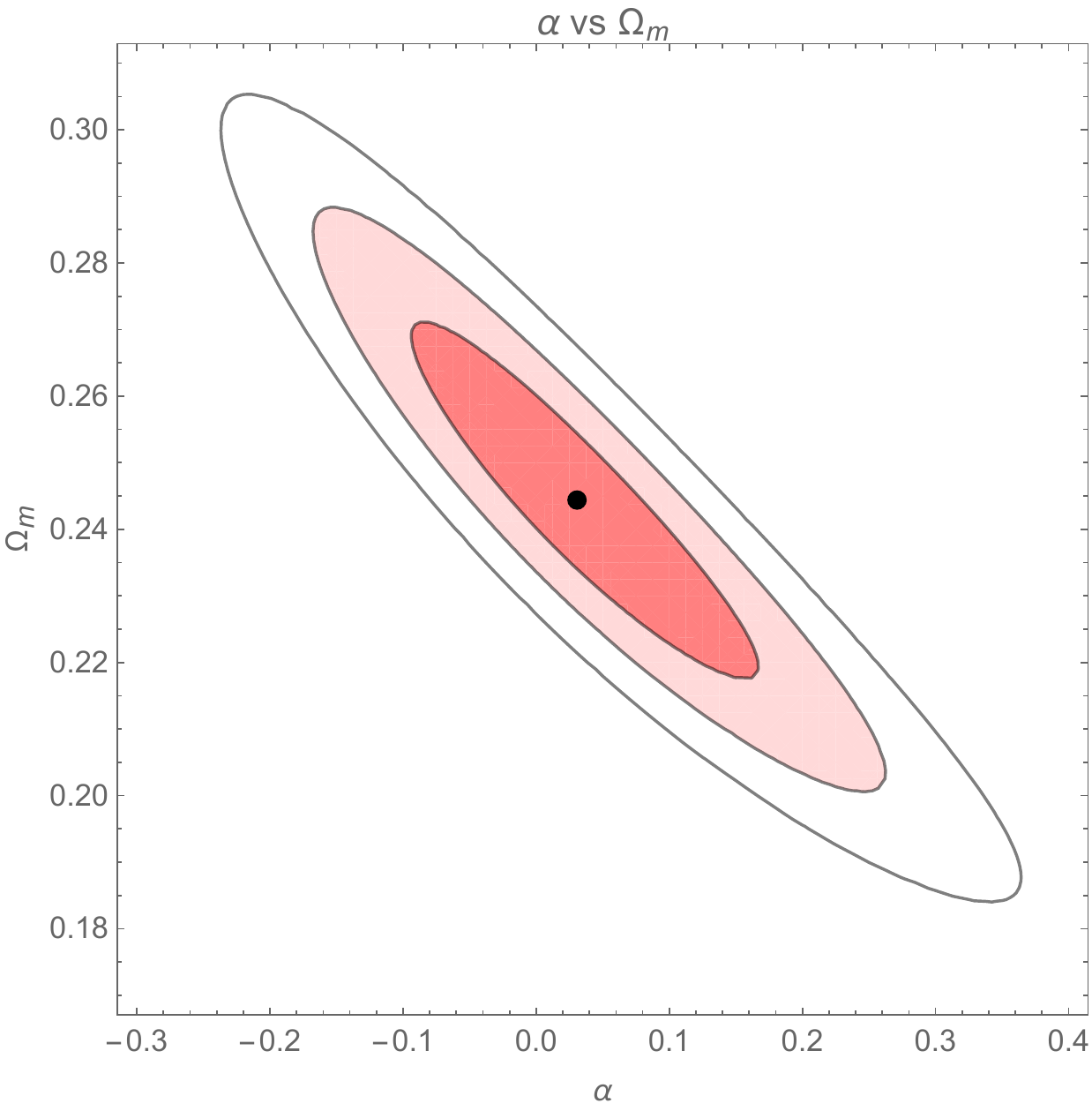}%

        } &
        \parbox{2in}{%
           \includegraphics[width=2in,height=1in]{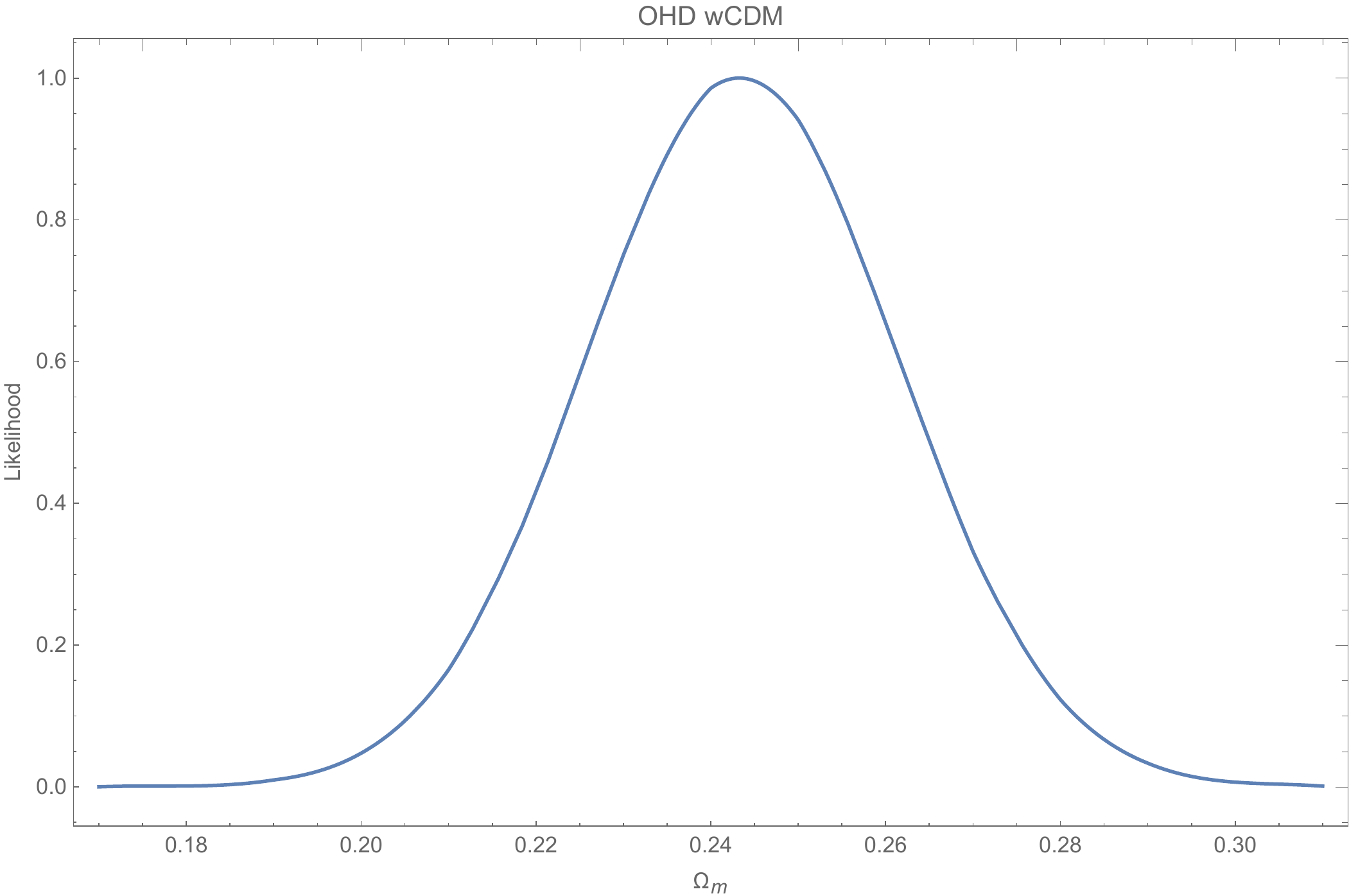}\\
            \includegraphics[width=2in,height=1in]{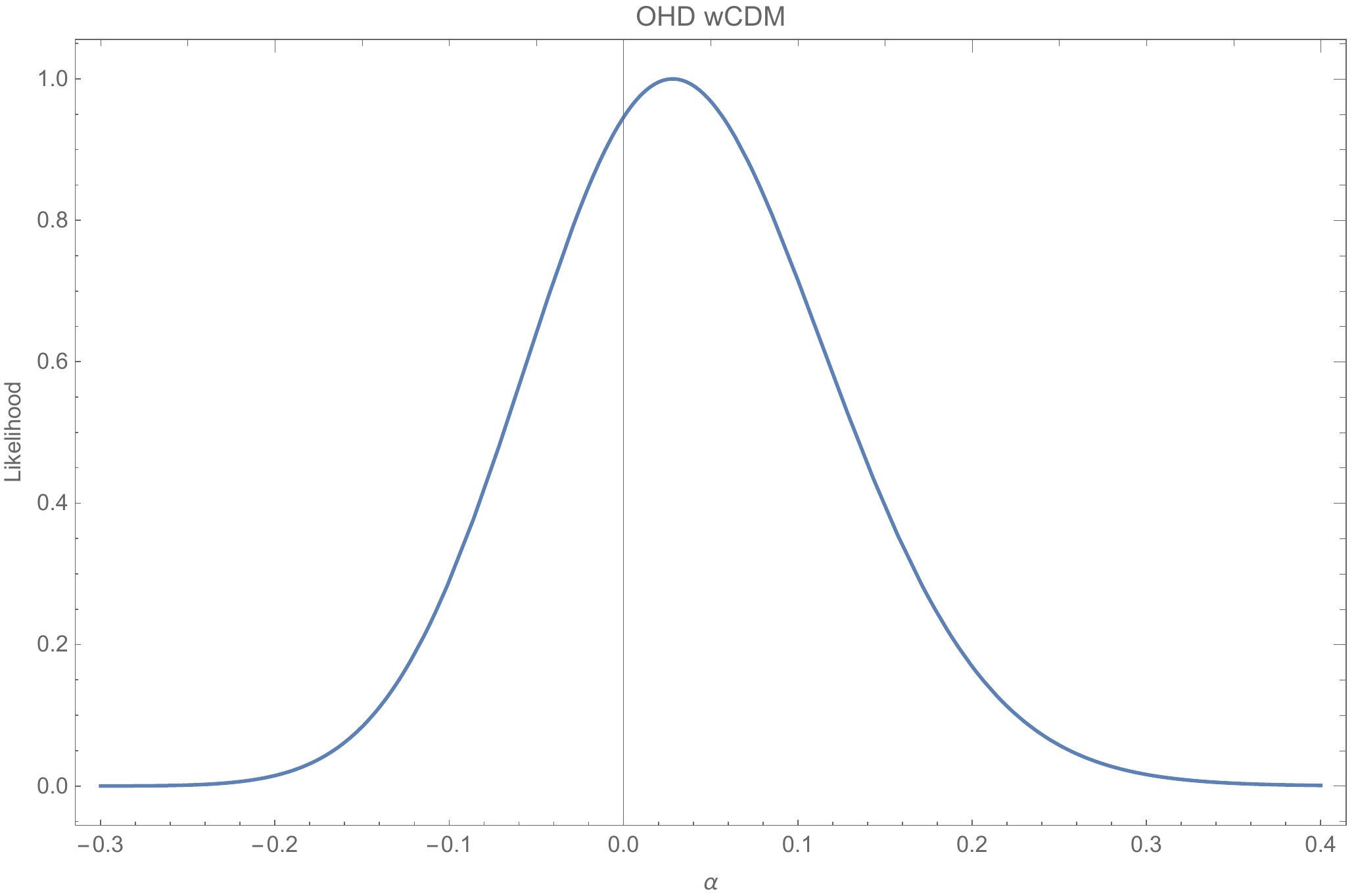}%
        } \\
    \end{tabular}
    \caption{\small\emph{$\Omega_m$ vs $\alpha$ graph with liklihood}} 
    \label{Comega}
\end{figure}
From contour the graph (see fig.~\ref{Comega}) we determine the value of $\alpha$ and $\Omega_m$  for $ H_0 = 71 ~km s^{-1}Mpc^{-1}$~\cite{sah} as presented  in the Table-\ref{t0}. In a previous study, Malekjani \textit{et al.}~\cite{male} obtained a value for $\alpha$ that is nearly identical, specifically $\alpha =0.033$.
\begin{table}[h!]
\centering
\begin{tabular}{|c|c|c|}
\hline
$\chi^2_{m}$ & $\alpha$ & $\Omega_m$ \\ \hline
$44.77$ & $0.03$ & $0.2443$ \\ \hline
\end{tabular}
\caption{\small\emph{Best-fit values of $\alpha$ and $\Omega_m$.}}
\label{t0}
\end{table}
Now the range of  $\alpha$ and $\Omega_m$ obtained by using the Hubble dataset and the results in the $1 \sigma$  confidence region is shown in Table-\ref{t01} .

\begin{table}[h!]
\centering
\begin{tabular}{|c|c|c|}
\hline
Range & $\alpha$ & $\Omega_m$  \\ \hline
$1 \sigma$  region& $-0.094,~ 0.167$ & $0.218,~ 0.271$  \\ \hline
\end{tabular}
\caption{\small\emph{The range of $\alpha$ and $\Omega_m$ in $1 \sigma$ region.}}
\label{t01}
\end{table}
We consider only positive values of $\alpha$ since we are dealing with the Generalized Chaplygin Gas (GCG) model, where $\alpha > 0$ is a fundamental parameter controlling the equation of state. Furthermore, our analysis indicates that $\alpha$ is not only small but also satisfies $\alpha < 1$, which is in agreement with our previous findings. This condition aligns exceptionally well with observational data, which strongly favor small values of $\alpha$ for consistency with cosmological measurements.

It is worth noting that this constraint on $\alpha$ is in direct contrast to the pure Chaplygin Gas model, which corresponds to $\alpha = 1$. The GCG model's flexibility, especially with $\alpha < 1$, allows for a smoother and more realistic transition from matter-dominated behavior to dark energy-like behavior, making it more consistent with observations.\\

Now the present age of the universe is given by
\begin{equation}\label{eq:20d}
t_0 = \int_{0}^{\infty} \frac{1}{(1+z)H(z)} dz
\end{equation}

Using the parameter values from Table-\ref{t0} and eq.~\eqref{eq:21b}, we obtain $t_0=13.95$ Gyr. This value is slightly higher than the result obtained from the Planck 2020 data, which estimates the age of the universe to be $13.8 $ Gyr \cite{pla}.

It is important to mention that the present age of the universe here has been calculated within the framework of the Generalized Chaplygin Gas (GCG) model using the Hubble $57$ dataset, which consists of direct measurements of the Hubble parameter $ H(z)$ at various redshifts in the late-time universe. In contrast, the Planck 2020 results are derived from observations of the cosmic microwave background at  $z \sim 1100$~\cite{na}.
This slight discrepancy can be attributed to the fact that these datasets probe different epochs and aspects of cosmic evolution. Additionally, variations in the assumed values of the Hubble constant and other cosmological parameters across different analyses can naturally lead to minor differences in the inferred age of the universe.

As pointed earlier  the key eq.~\eqref{eq:10} is not amenable to  an explicit solution  which is a function of time in known simple form. In this case the variation of cosmological variables like sale factor, flip time  etc. can not be explicitly obtained. To avoid such a difficulty of obtaining solution in known form to  determine
the flip time and other physical features of cosmology we adopt here an \emph{alternative approach}~\cite{dp3, dp4} in the next section.

\vspace{0.2 cm}

\section{ An alternative approach :}

 \vspace{0.2 cm}
As we consider the late evolution of our model, the second term on the right-hand side (RHS) of eq.~\eqref{eq:10} becomes almost negligible compared to the first term. The Chaplygin gas equation of state describes the transition from a dust-dominated universe to the present accelerating phase. Therefore, the scale factor should be sufficiently large, and the ratio of the model parameters is small in this case. Under these conditions, it is reasonable to consider only the first-order approximation of the binomial expansion of the RHS of eq.~\eqref{eq:10}. In this work, we derive an exact solution using the first-order approximation of eq.~\eqref{eq:10}, building upon similar attempts made in earlier works~\cite{dp3, dp4}. Simplifying eq.~\eqref{eq:10} by neglecting higher-order terms, the equation for the late stage of evolution reduces to

\begin{equation}\label{eq:17}
3 \frac{\dot{a}^2}{a^2} =   B^{\frac{1}{1+\alpha}} + \frac{c}{(1+\alpha) B^{\frac{\alpha}{1+\alpha}}}a^{-3(1+\alpha)}
 \end{equation}

\begin{figure}[ht]
\begin{center}
  \includegraphics[width=8 cm]{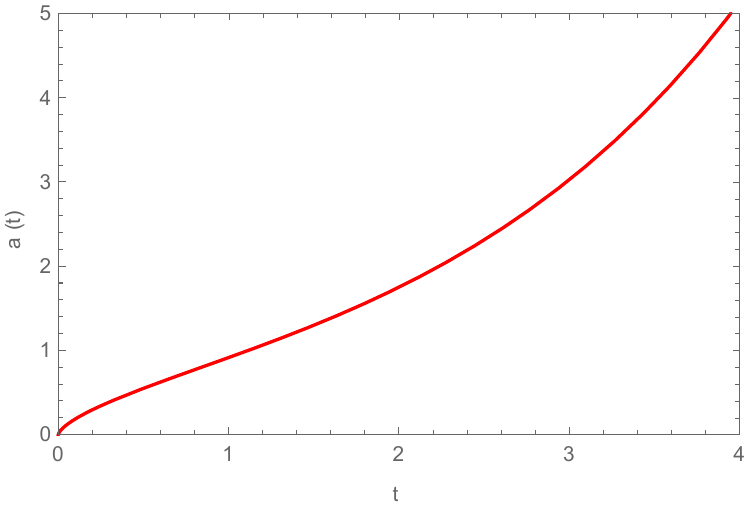}
  \caption{
  \small\emph{The variation of $a$ and $t$    }\label{at}
    }
\end{center}
\end{figure}

Solving the eq.~\eqref{eq:17} we get an explicit  solution of the scale factor as

\begin{equation}\label{eq:18}
a(t) = a_{0} \sinh^{n} \omega t
 \end{equation}

where, $a_{0} =  \left \{ \frac{1}{(1+\alpha)} \right \}^{\frac{1}{3(1+\alpha)}}$ ; $n = \frac{2}{3(1+\alpha)}$ and $\omega = \frac{\sqrt{3}}{2} (1+\alpha) B^{\frac{1}{2(1+\alpha)}}$.

In fig.-\ref{at}, we observe the evolution of the scale factor $a$ with respect to time $t$.

\begin{figure}[ht]
\begin{center}
  \includegraphics[width=10cm]{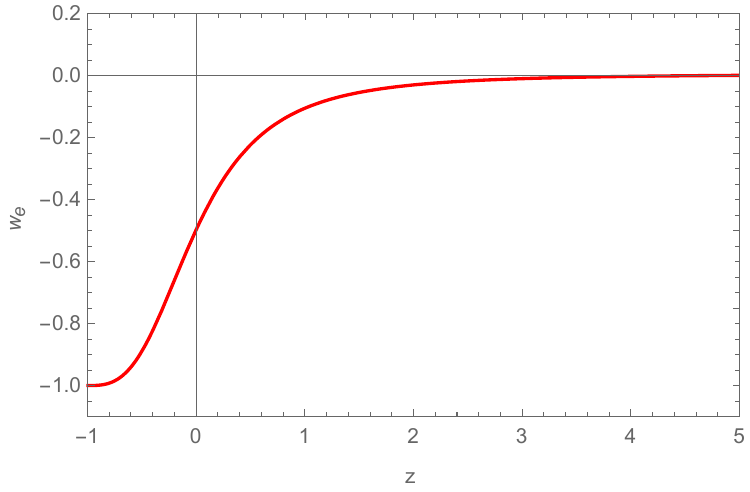}
  \caption{
  \small\emph{The variation of $w$ with $z$ is shown in this figure.  }\label{wza}
    }
\end{center}
\end{figure}

Now using eqs~\eqref{eq:3}, \eqref{eq:4} and \eqref{eq:18} we can write the expressions of $\rho$ and $p$ as follows.

\begin{equation}\label{eq:18a}
\rho = 3n^2 \omega^2 \coth^2\omega t = B^{\frac{1}{1+\alpha}} \left\{1+(1+z)^{3(1+\alpha)}\right\}
 \end{equation}

and

\begin{equation}\label{eq:18b}
p = n \omega^2 \left\{(2-3n) \coth^2 \omega t -2 \right \} = - B^{\frac{1}{1+\alpha}} \left\{1-\alpha(1+z)^{3(1+\alpha)}\right\}
 \end{equation}
\begin{figure}[ht]
\begin{center}
  \includegraphics[width=8 cm]{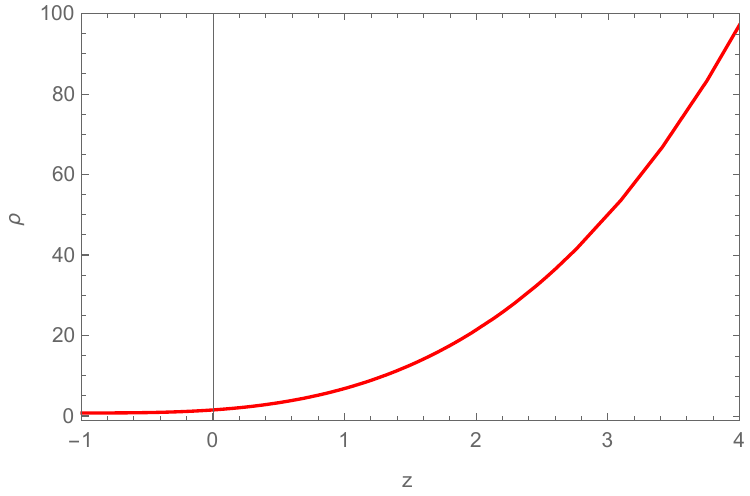}
  \caption{
  \small\emph{The variation of $\rho$ and $z$    }\label{rza}
    }
\end{center}
\end{figure}
Fig.-\ref{rza} shows that the matter density $\rho$ decreases with the redshift parameter $z$, as expected.
Now the effective equation of state is given by

\begin{equation}\label{eq:18b}
w_e = \frac{p}{\rho} = \frac{2-3n}{n^2} - \frac{2}{3} \tanh^2 \omega t = - \frac{1-\alpha(1+z)^{3(1+\alpha)}}{1+(1+z)^{3(1+\alpha)}}
 \end{equation}
The eq.~\eqref{eq:18b} gives the following results:
\begin{itemize}
    \item[(i)] In a dust-dominated universe, where $w_e = 0$, the redshift is given by $z = \frac{1}{\alpha^{\frac{n}{2}}} - 1$. For this expression to yield a positive $z$, we require $\alpha < 1$. Furthermore, for a pure Chaplygin gas with $\alpha = 1$, we obtain $z = 0$. This implies that, when observed from a dust-dominated timescale, the universe appears as if it were filled with a pure Chaplygin gas.

    \item[(ii)] At the present epoch (\textit{i.e.}, at $z = 0$), we find that $w_e = - 1 + \alpha$, indicating an accelerating universe and in our case, $w_e = -0.9949$. Interestingly for a pure Chaplygin gas ($\alpha = 1$), this result resembles  a dust-dominated universe ($w_e = 0$), which is consistent with earlier discussions.

    \item[(iii)] In the later stages of the evolution of the universe, we find $w_e = -1$, corresponding to a $\Lambda$CDM model, as detailed in Section-3.2.
\end{itemize}
It is observed that $w_e$ gradually transitions from 0 to $-1$, representing the evolution of the generalized Chaplygin gas  from a dust-dominated phase to a $\Lambda$CDM-type cosmology (see fig.~\ref{wza}). Additionally, it is noted that $\alpha$ becomes zero for $n = \frac{2}{3}$, where $p = -B$. It is worth mentioning the exact analytical form of the scale factor at late times for the standard $\Lambda$CDM model, where the exponent is $n = \frac{2}{3}$.

From eq.~\eqref{eq:18}  we get the deceleration parameter

\begin{equation}\label{eq:19}
q = \frac{1-n \cosh^2 \omega t}{n \cosh^2 \omega t}= \frac{1}{n\{1+(1+z)^{-\frac{2}{n}}\}}-1
 \end{equation}
The eq.~\eqref{eq:19} indicates that the exponent $n$ in eq.~\eqref{eq:18} governs the evolution of $q$.
 A little analysis of eq.~\eqref{eq:19} shows that (i) if  $n > 1$ we get only  acceleration, no \emph{flip} occurs in this condition. But for $n > 1$ gives $- \frac{1}{3} > \alpha$, which is physically unrealistic, since  we know that  $\alpha > 0$. (ii) Again, if $0>n>\frac{2}{3}$ it gives early deceleration and late acceleration and in this condition $\alpha > 0$, so the desirable feature of \emph{flip} occurs which agrees with the observational analysis for positive values  of $\alpha$.

The analysis of $q$ using eq.~\eqref{eq:19} reveals the following:

\begin{itemize}
    \item[(i)] In the early universe, \textit{i.e.} at high $z$, $ q = \frac{1}{n} - 1 $, this corresponds to a dust-dominated universe. In this case, $ n = \frac{2}{3(1+\alpha)} $, and $ q \approx 0.5 $ (from Table-\ref{t120}), which is in very good agreement with  present well-known results. We obtained the same result in Section-3.1;

    \item[(ii)] At the present epoch (\textit{i.e.}, $ z = 0 $), eq.~\eqref{eq:19} reduces to $ q = \frac{1}{2n} - 1 $. we get the value of  $ q \approx  - 0.996 $ (from Table-\ref{t120}), which corresponds to an accelerating universe.

    \item[(iii)] In the late universe, we find $ q = -1 $, which represents a pure $\Lambda$CDM model. It is notable that $q$  approaches $-1$  from the past towards the present epoch, consistent with a transition to acceleration over cosmic time.
\end{itemize}

\begin{figure}[ht]
\begin{center}
  \includegraphics[width=10cm]{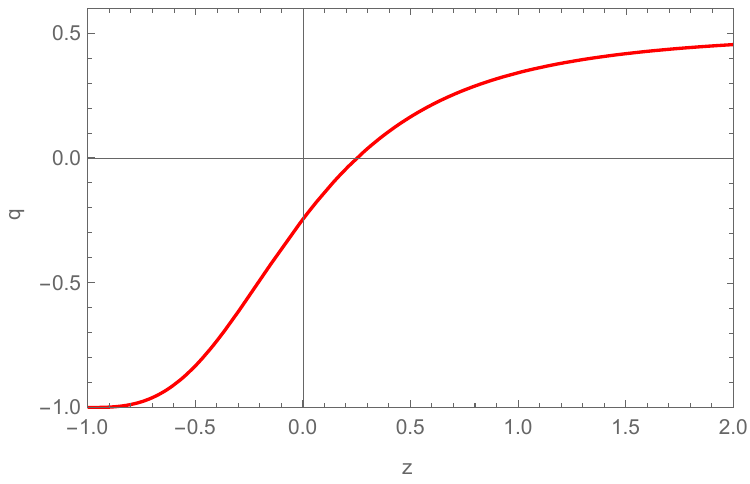}
  \caption{
  \small\emph{The variation of $q$ with $z$ is shown in this figure.    }\label{qza}
    }
\end{center}
\end{figure}
Fig.-\ref{qza}  shows the variation of $q$ with $z$. We would like to focus on the occurrence of late flip as because all observational evidences suggest that accelerating phase is a recent phenomena. It is interesting to note that the late flip also depends on the value of $\alpha$.   Now the \emph{flip} time $(t_f)$ will be in this case

\begin{equation}\label{eq:20}
t_f = \frac{1}{\omega} \cosh^{-1} \left(\sqrt{\frac{1}{n}} \right) = \frac{1}{\omega} \cosh^{-1}\sqrt{\frac{3(1+\alpha)}{2}}
\end{equation}

\begin{figure}[ht]
\begin{center}
  \includegraphics[width=10cm]{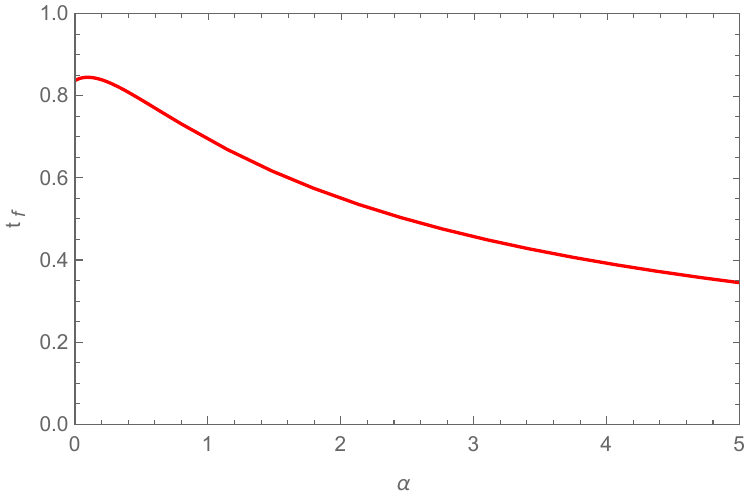}
  \caption{
  \small\emph{The graphs clearly show that flip time depends on  $\alpha$.   }\label{tfa}
    }
\end{center}
\end{figure}
Using eq.~\eqref{eq:20} we have drawn the fig.-\ref{tfa} where the variation of $t_f$ with $\alpha$ is shown. When the value of $\alpha$ is small $t_{f}$ increases, i.e., small $\alpha$ favours the late flip. From the observation, we get the small value of $\alpha = 0.0051$.
Now from eq.~\eqref{eq:19} the redshift parameter at flip is given by,
\begin{equation}\label{eq:28a}
z_f = \left(\frac{n}{1-n} \right)^{\frac{n}{2}} -1
\end{equation}
This also implies $n < 1$ for $z_f$
  to be real, a condition that aligns with observational findings. To achieve acceleration at the present epoch, we need $z_f >0$, which further imposes the constraint on  $\alpha < \frac{1}{3}$. For acceleration at the present epoch, we find that $\alpha < \frac{1}{3}$ (in our case $\alpha = 0.0051$), a condition that is consistent with values obtained from observational data. \\
Now, using eqs.~\eqref{eq:18c} and \eqref{eq:19}, we obtain the expression for the jerk parameter as

\begin{dmath}\label{eq:29}
j = \left[\frac{1}{n \left \{1 + (1+z)^{-\frac{2}{n}} \right \} }- 1 \right]\left[\frac{2}{n \left \{1 + (1+z)^{-\frac{2}{n}} \right \} }- 1 \right] \\
 + \left[\frac{2(1+z)^{-\frac{2}{n}}}{n^2 \left \{1 + (1+z)^{-\frac{2}{n}} \right \}^{2} }  \right]
\end{dmath}
\par
\begin{figure}[ht]
\begin{center}
  \includegraphics[width=10cm]{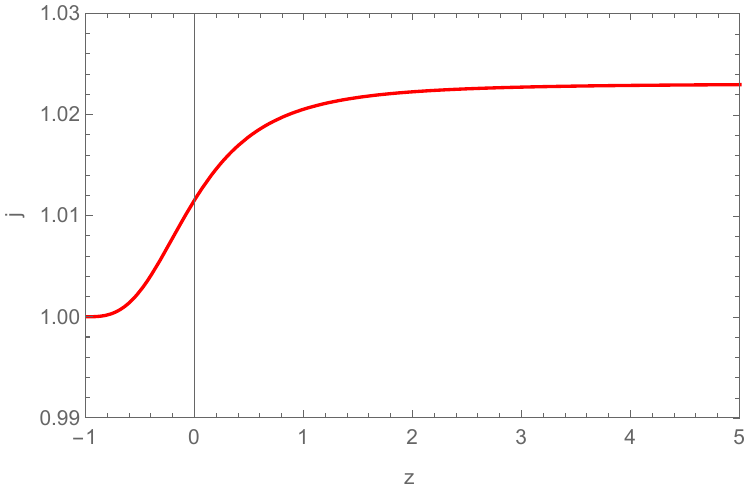}
  \caption{\small\emph{The evolution of the jerk parameter $j$ as a function of redshift $z$. The graph clearly shows that $j$ approaches unity at late times, corresponding to the $\Lambda$CDM model. } \label{jza}}
\end{center}
\end{figure}

The behavior of the jerk parameter in different cosmological epochs is summarized as follows:
\begin{itemize}
    \item[(i)] In the early universe, \textit{i.e.}, at high $z$, eq.~\eqref{eq:29} reduces to $ j = \frac{1}{2}(1 + 3\alpha)(2 + 3\alpha)$,     which corresponds to a dust-dominated phase. Substituting the value of $\alpha = 0.0051$, we obtain $j \approx 1.02307$ which is slightly higher than unity, signifying a dust-like behavior with a minor GCG correction in the early phase, as shown in fig.-\ref{jza}.

    \item[(ii)] At the present epoch (\textit{i.e.}, $z = 0$), eq.~\eqref{eq:29} simplifies to $ j = \frac{1}{4}(9 \alpha^2 + 9 \alpha + 4)$, yielding $j \approx 1.01153$. The slight departure from $j=1$ signifies that although the universe’s expansion closely follows the $\Lambda$CDM model, it retains a mild signature of the GCG’s unifying dark sector dynamics, which begin to manifest more noticeably in the current and future epochs.

    \item[(iii)]  In the late-time universe (\textit{i.e.}, at large cosmic time), the model asymptotically approaches $j = 1$, which corresponds to a pure $\Lambda$CDM-like behavior, as also depicted in fig.~\ref{jza}.

\end{itemize}
Therefore, the Generalized Chaplygin Gas (GCG) model can serve as a unifying fluid for the dark sector, effectively reproducing the cosmic expansion history in close agreement with the standard $\Lambda$CDM  model within observational constraints. However, it provides a subtly different physical framework that may offer new insights. Using eq.~\eqref{eq:29}, the graphical representation of the jerk parameter is presented in fig.~\ref{jza}.

As evident from fig.-\ref{jza}, the value of $j$ asymptotically approaches $1$, which is a characteristic feature of the $\Lambda$CDM cosmological model at future cosmic times. This is in good agreement with current observational data regarding the dynamics of our universe.

Now the expression of Hubble parameter in terms of $\Omega_m (= \frac{c}{\rho_0^{1+\alpha}})$ is given by

\begin{equation}\label{eq:30}
H = H_0 \Omega_m^{\frac{1}{2(1+\alpha)}}\left \{\left(\frac{1 - \Omega_m}{\Omega_m}\right)^{\frac{1}{1+\alpha}}  + \frac{1}{1 + \alpha}\left(\frac{\Omega_m}{1 - \Omega_m}\right)^{\frac{\alpha}{1+\alpha}}
(1+z)^{3(1+\alpha)}\right\}^{\frac{1}{2}}
\end{equation}

\begin{figure}[h!]
    \centering 
    \begin{tabular}{l l}
        \parbox{2in}{%
            \includegraphics[width=2in,height=2.1in]{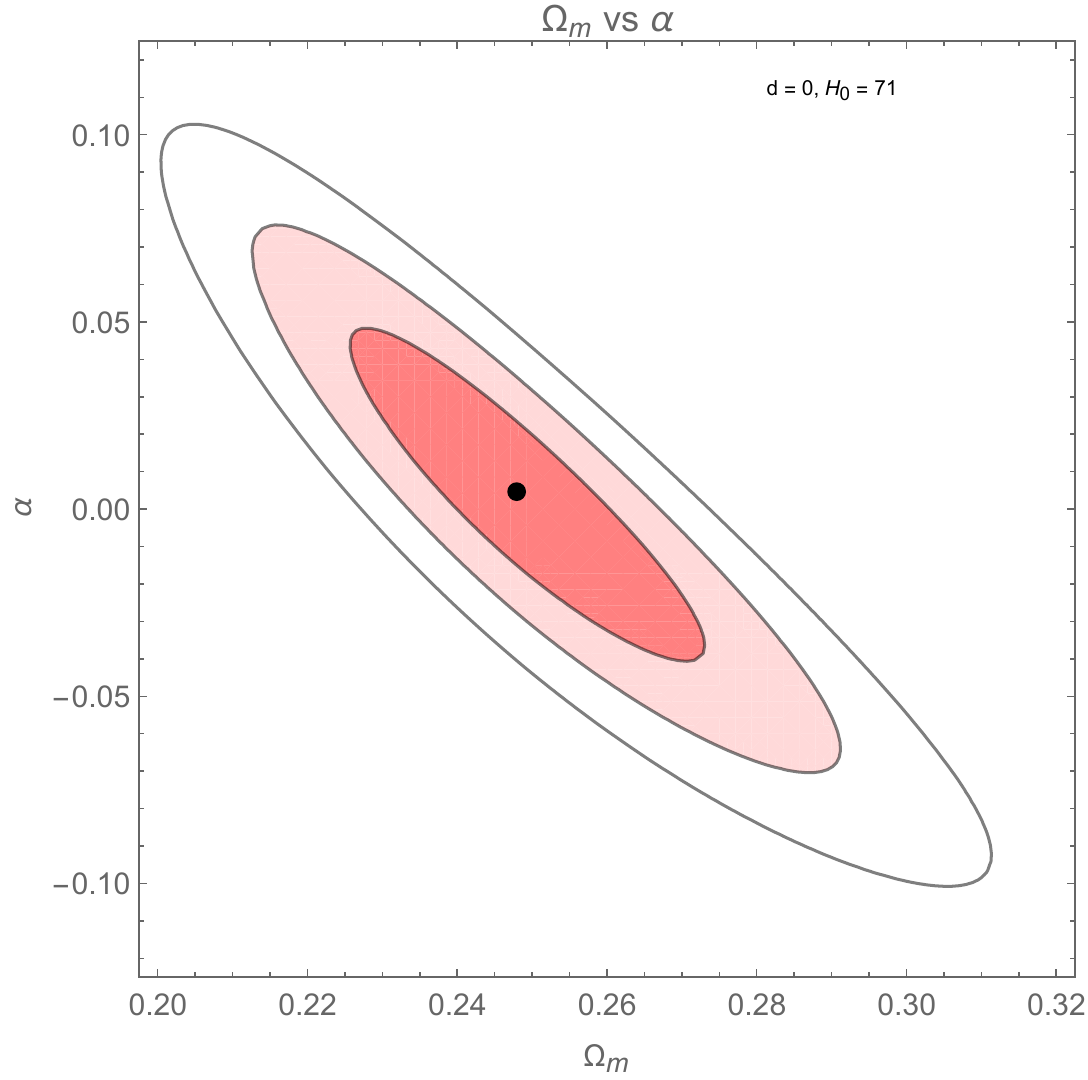}%

        } &
        \parbox{2in}{%
           \includegraphics[width=2in,height=1in]{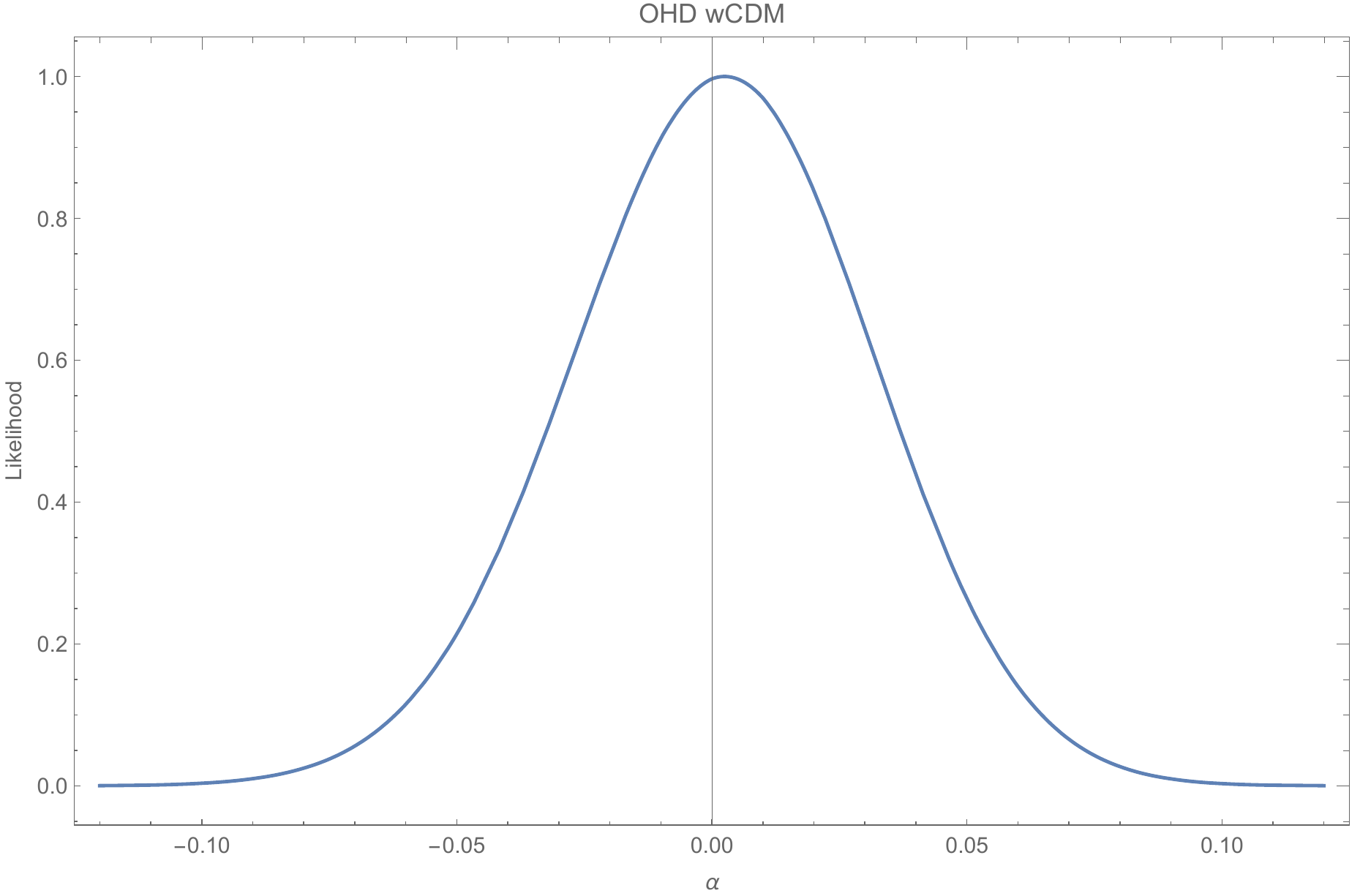}\\
            \includegraphics[width=2in,height=1in]{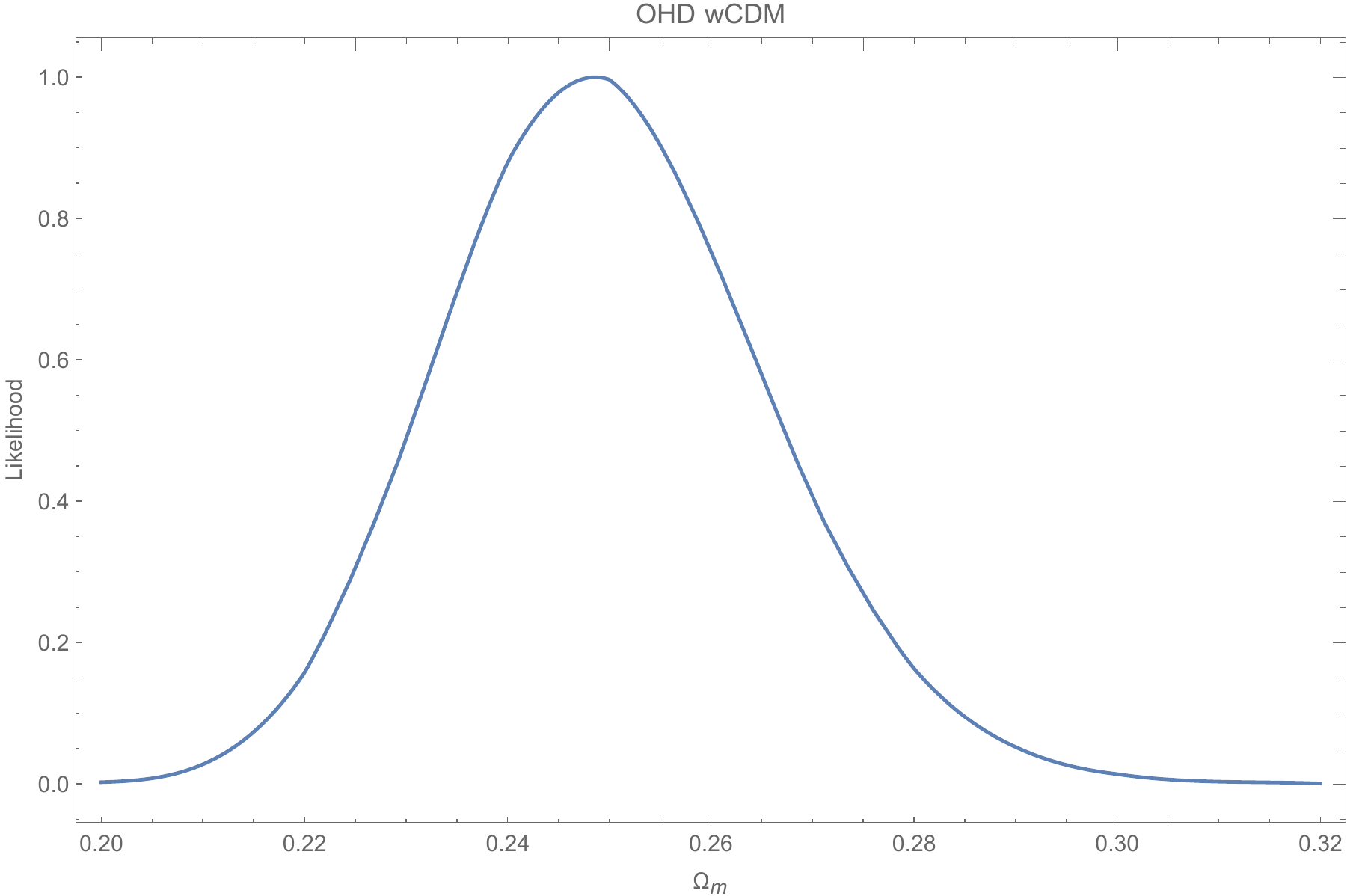}%
        } \\
    \end{tabular}
    \caption{\small\emph{$\Omega_m$ vs $\alpha$ graph with liklihood} }
    \label{Coaa1}
\end{figure}

A contour plot is drawn as shown in fig.-\ref{Coaa1} using eq.~\eqref{eq:30}. The best-fit values of $\alpha$ and $\Omega_m$ are given in Table-\ref{t120}, and the ranges within the $1 \sigma$ confidence region are presented in Table-\ref{t221}.

\begin{table}[h!]
\centering
\begin{tabular}{|c|c|c|}
\hline
$\chi^2_{m}$ & $\Omega_m$ & $\alpha$ \\ \hline
$44.87$  & $0.248$ & $0.0045$  \\ \hline
\end{tabular}
\caption{\small\emph{Best-fit values of $\Omega_m$ and $\alpha$. }}
\label{t120}
\end{table}

\begin{table}[h!]
\centering
\begin{tabular}{|c|c|c|}
\hline
Range & $\Omega_m$ & $\alpha$  \\ \hline
$1~\sigma$ region & $0.2262,~ 0.2719$ & $-0.0410,~ 0.0470$  \\ \hline
\end{tabular}
\caption{\small\emph{The range of $\alpha$ and $\Omega_m$ in $1 \sigma$ region.}}
\label{t221}
\end{table}
From eqs.~\eqref{eq:20d} \& ~\eqref{eq:30}, and using the parameter values from Table-\ref{t120}, we determine the present age of the universe as $t_0 = 13.51$ Gyr. This value is slightly lower than the result obtaining from Planck 2020 data~\cite{pla}.

\begin{figure}[h!]
    \centering 
    \begin{tabular}{l l}
        \parbox{2in}{%
            \includegraphics[width=2in,height=2.1in]{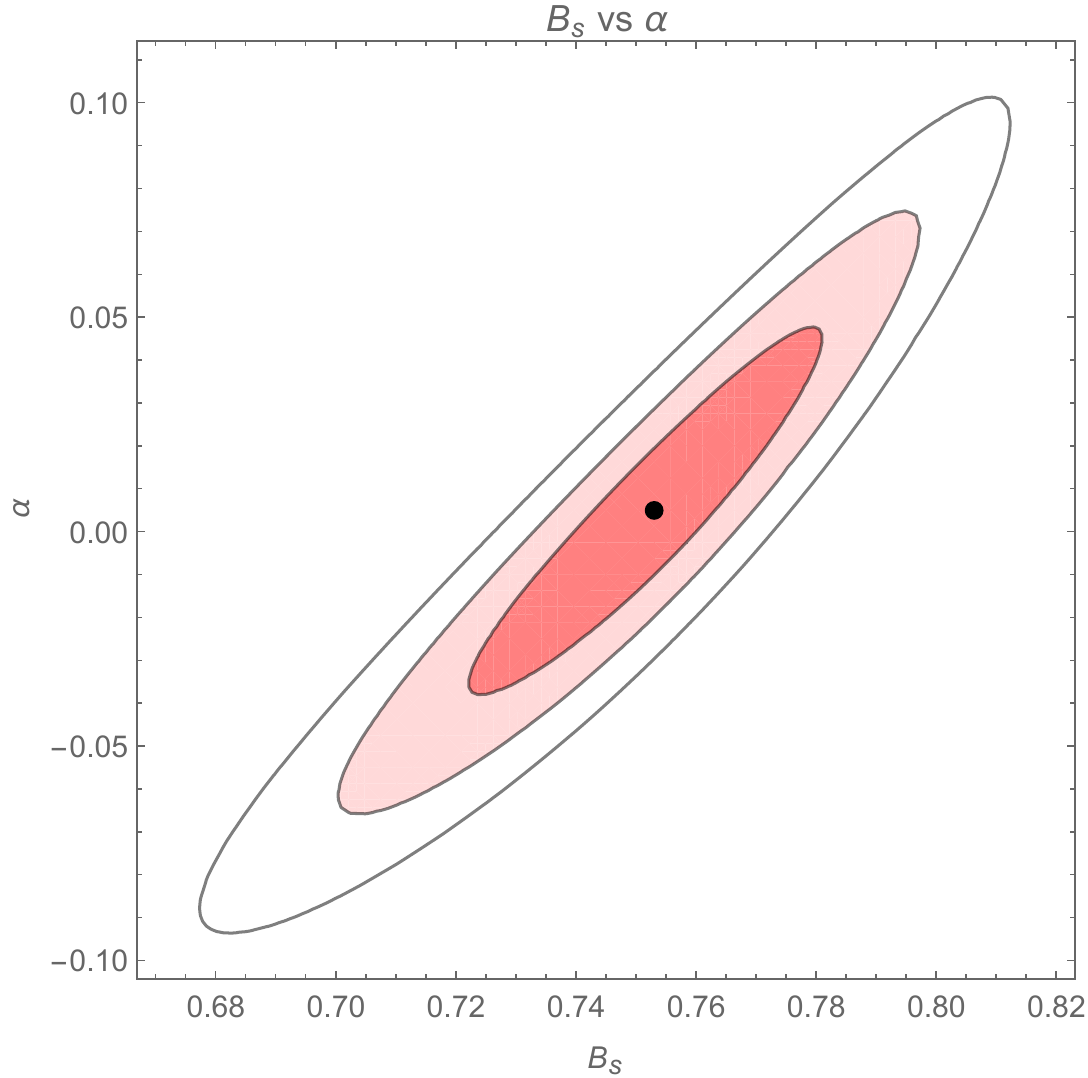}%

        } &
        \parbox{2in}{%
           \includegraphics[width=2in,height=1in]{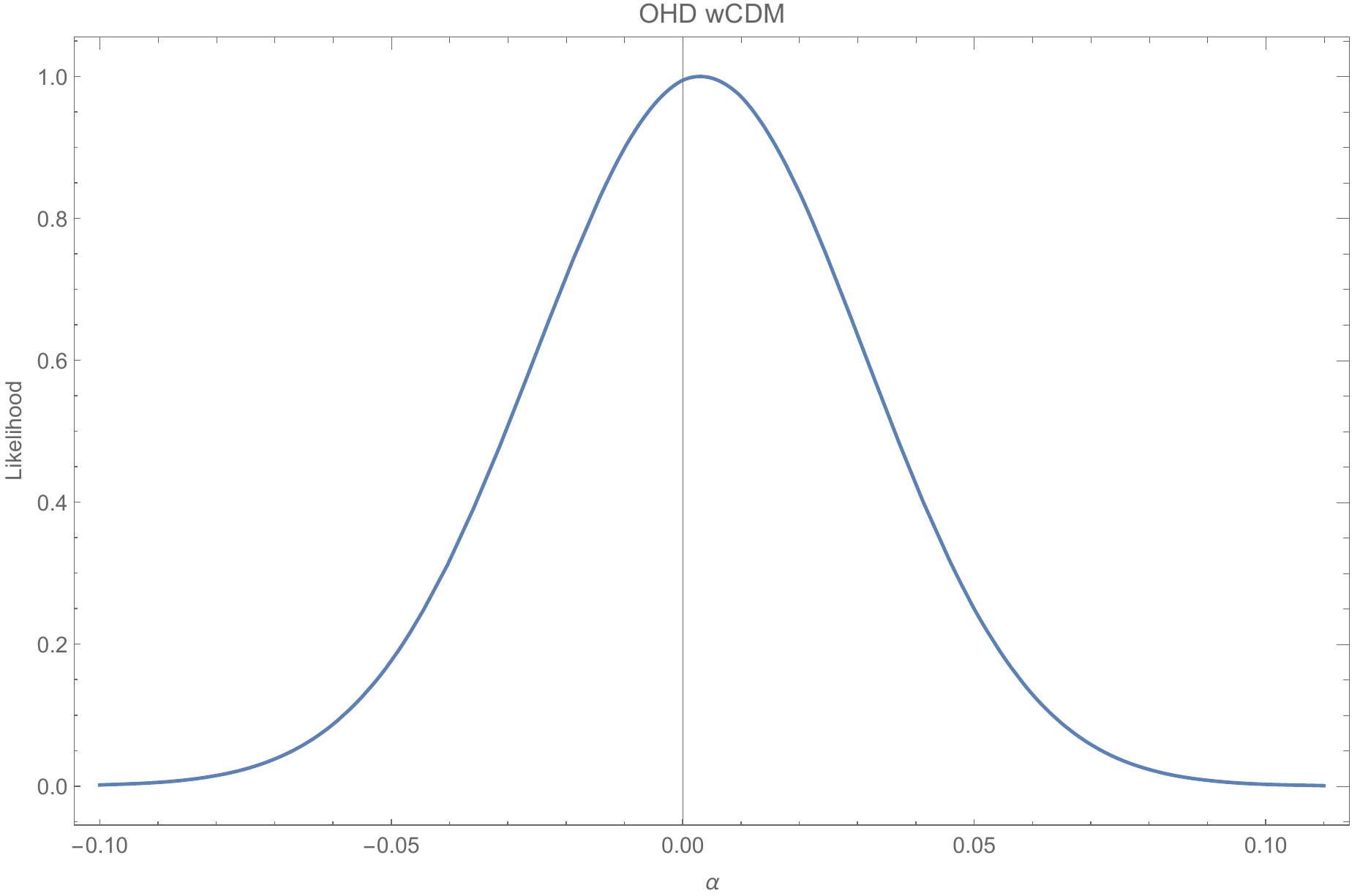}\\
            \includegraphics[width=2in,height=1in]{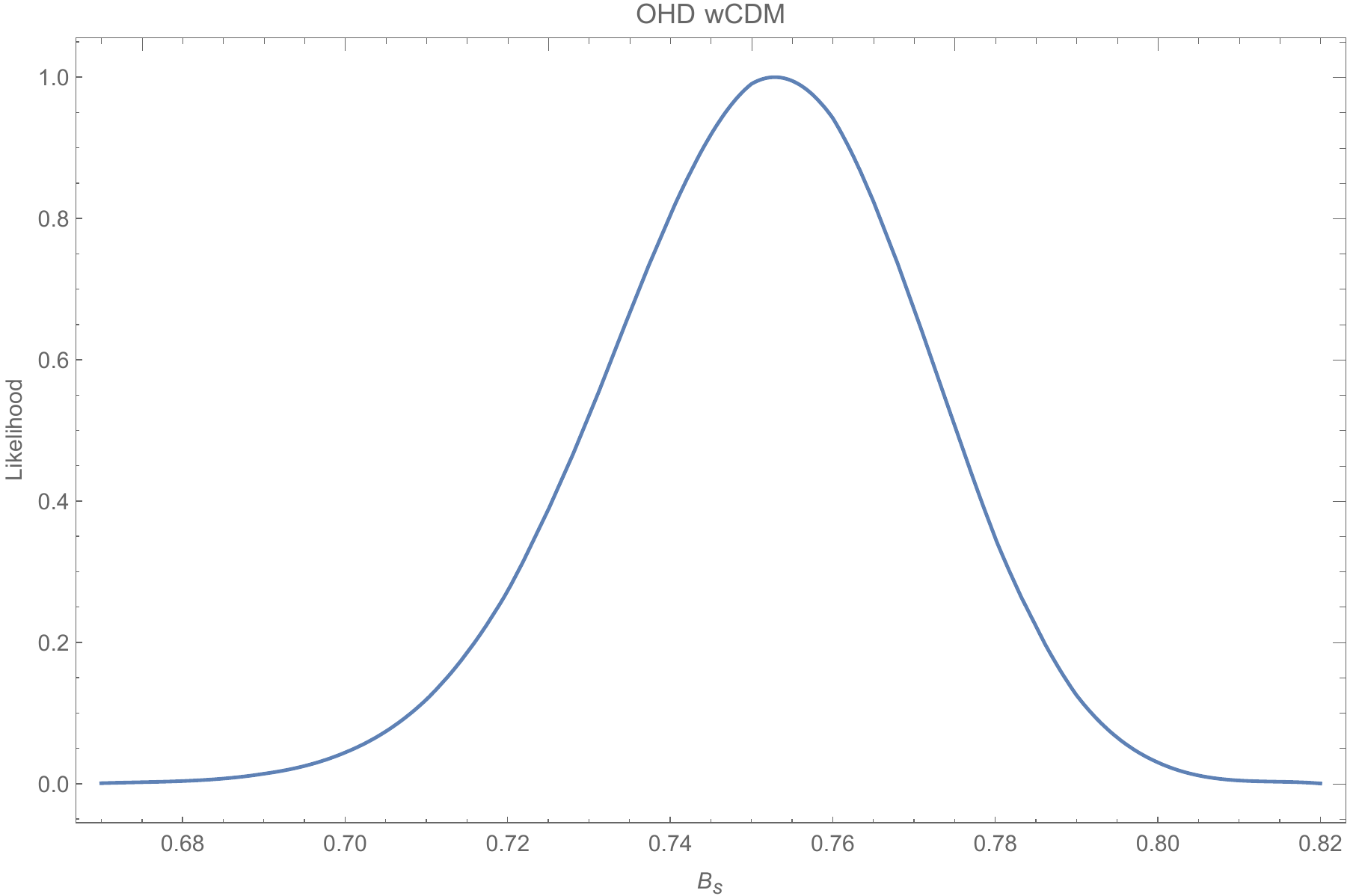}%
        } \\
    \end{tabular}
    \caption{\small\emph{$B_s$ vs $\alpha$ graph with liklihood}} 
    \label{Lbsa}
\end{figure}

To constrain the parameters, let us consider $B_s = \frac{B^{\frac{1}{1+\alpha}}}{\rho_0}$, we get the expression of Hubble parameter in terms of $B_s$ as

\begin{equation}\label{eq:31}
H = H_0 \left \{ B_s  +\left(1- B_s \right)\left(1+z \right)^{3(1+\alpha)}
\right\}^{\frac{1}{2}}
\end{equation}
The values of $B_s$ and $\alpha$ at $\chi^2_{m}$ are determined from the contour graph (see fig.-\ref{Lbsa}) in Table~\ref{t20} and the ranges in Table-\ref{t21}. We observe that $\alpha < 1$, contrasting with the pure Chaplygin gas model where $\alpha = 1$. The Generalized Chaplygin Gas (GCG) model has been extensively studied by several authors, who have constrained the model parameters using various observational data. For instance, Malekjani \textit{et al.}~\cite{male} obtained a value of $B_s = 0.76$, closely aligning with our findings. Similarly, Bertolami et al.~\cite{bert} determined that $B_s$ ranges from $0.62$ to $0.82$, and $\alpha$ ranges from $0.052$ to $1.056$; P. Wu and H. Yu~\cite{wu} found that $0.67 \leq B_s \leq 0.83$ and $-0.21 \leq \alpha \leq 0.42$. Additionally, P. Thakur~\cite{thak} obtained $B_s = 0.772$ and $\alpha = 0.023$. The values of $B_s$ and $\alpha$ found by these authors are comparable to those obtained through our alternative approach.
It is to be mentioned that the condition derived from the first approximation $\frac{B_s^{1+\alpha}}{\Omega_m}> \frac{1}{2}$, is also satisfied by the value obtained from our alternative approach, which confirms the consistency of the results.

\begin{table}[h!]
\centering
\begin{tabular}{|c|c|c|c|}
\hline
$\chi^2_{m}$ & $B_s$ & $\alpha$ \\ \hline
$44.861$ & $0.7532$ & $0.0051$  \\ \hline
\end{tabular}
\caption{\small\emph{Best-fit values of $B_s$ and $\alpha$.}}
\label{t20}
\end{table}

\begin{table}[h!]
\centering
\begin{tabular}{|c|c|c|}
\hline
Region & $B_s$ & $\alpha$  \\ \hline
$1~\sigma$ & $0.7226,~ 0.7807$ & $-0.0372,~ 0.0475$  \\ \hline
\end{tabular}
\caption{\small\emph{The range of $\alpha$ and $B_s$ in $1 \sigma$ region.} }
\label{t21}
\end{table}
We also determine the present age of the universe using the parameter values from Table~\ref{t20}. From eqs.~\eqref{eq:20d} and \eqref{eq:31}, we calculate $t_0 = 13.52$ Gyr.  This value is slightly higher than the result obtained from Planck 2020 data~\cite{pla}. This slight discrepancy may be attributed to the fact that these datasets probe different epochs and aspects of cosmic evolution. Moreover, variations in the assumed values of the Hubble constant and other cosmological parameters across different analyses naturally lead to minor differences in the estimated age of the universe.

\section{Raychaudhuri Equation}

It may not be out of place to address and compare the situation
discussed in the last section with the help of the well known Raychaudhuri equation~\cite{ray},  which in general holds for any
cosmological solution based on Einstein's gravitational field
equations. With matter field expressed in terms of mass density
and pressure Raychaudhuri equation  reduces to a compact form as

\begin{equation}\label{eq:32}
  \dot{\theta}=-2(\sigma^{2}-\omega^{2})-\frac{1}{3}\theta^{2}-\frac{8\pi G}
  {2}\left(\rho+3p \right)
\end{equation}
in a  co moving reference frame. Here $p$ is the
 isotropic pressure and $\rho$ is the energy density from varied sources.
\vspace{0.1 cm} Moreover other quantities are defined with the
help of a unit vector $v^{\mu}$ as under

\begin{eqnarray}\label{eq:33}\nonumber
  \textrm{the expansion scalar}~~\theta & = & v^{i};_{i} \\ \nonumber
 \sigma^{2} & = & \sigma_{ij}\sigma^{ij}  \\ \nonumber
\textrm{the shear tensor} ~\sigma_{ij} &=& \frac{1}{2}(v_{i;j}+
v_{j;i})- \frac{1}{2}(\dot{v}_{i}v_{j}+\dot{v}_{j}v_{i})-
\frac{1}{3}v^{\alpha}_{;\alpha}(g_{ij} - v_{i}v_{j})  \\
\textrm{the vorticity tensor}~~  \omega_{ij} & = &
\frac{1}{2}(v_{i;j}- v_{j;i})
-\frac{1}{2}(\dot{v}_{i}v_{j}-\dot{v}_{j}v_{i}
 \end{eqnarray}
We can calculate  an expression for effective
 deceleration parameter as
\begin{equation}\label{eq:34}
 q= -\frac{\dot{H} + H^{2}}{H^{2}} = -1-3~\frac{\dot{\theta}}{\theta^{2}}
\end{equation}
which allows us to write,
\begin{equation}\label{eq:35}
  \theta^{2}q = 6 \sigma^{2}+ 12 \pi G \left( \rho
  + 3p \right)
\end{equation}

\textbf{Case-1:} With the help of the eqs~\eqref{eq:7}, \eqref{eq:9} \& \eqref{eq:35} we finally get,

\begin{equation}\label{eq:36}
  \theta^{2}q =  6 \sigma^{2} + 12 \pi G
  \rho^{-\alpha}
  \left[-2B + \frac{c}{a^{3(1+\alpha)}}\right] = 6 \sigma^2 + 12\pi G \rho^{-\alpha} c \left[-\frac{2(1-\Omega_m)}{\Omega_m} +(1+z)^{3(1+\alpha)}\right]
\end{equation}
In our case as we are dealing with an isotropic rotation free
spacetime both  the shear scalar and vorticity vanish, i.e., $\sigma = 0$ and $\omega =0$, the eq.~\eqref{eq:36} now reduces to

\begin{equation}\label{eq:37}
  \theta^{2}q =  12 \pi G
  \rho^{-\alpha}
  \left[-2B + \frac{c}{a^{3(1+\alpha)}}\right]= 12\pi G \rho^{-\alpha} c \left[-\frac{2(1-\Omega_m)}{\Omega_m} +(1+z)^{3(1+\alpha)}\right]
\end{equation}

It follows from the eq.~\eqref{eq:37} that flip occurs (i.e., at $q=0$) when

\begin{equation}\label{eq:38}
a = \left(\frac{c}{2B} \right)^{\frac{1}{3(1+\alpha)}}
\end{equation}
and
\begin{equation}\label{eq:39}
z_f = \left\{\frac{2(1-\Omega_m)}{\Omega_m}\right\}^{\frac{1}{3(1+\alpha)}} -1
\end{equation}
Now $q < 0$, at $a
> \left(\frac{c}{2B} \right)^{\frac{1}{3(1+\alpha)}}$ \emph{i.e.}, acceleration takes place in this case.

It also follows from Raychaudhuri equation that our solution is in conformity with early deceleration and late acceleration. This result also agrees with the  eq.~\eqref{eq:19} for $\alpha >0$. It is interesting to note that the expression of scale factor at flip expressed by eqs~\eqref{eq:13a} and \eqref{eq:38} are identical.  Furthermore, the value of $z_f$ in eq.~\eqref{eq:39} agrees with our earlier findings in eq.~\eqref{eq:16b}.

\textbf{Case 2:} Now we have discussed our alternative approach in the context of Raychaudhuri equation where  shear scalar and vorticity vanish  because we have consider an isotropic rotation free spacetime. Now using eqs~\eqref{eq:18a}, \eqref{eq:18b} and \eqref{eq:35}  we finally get after straight forward calculation that

\begin{equation}\label{eq:40}
  \theta^{2}q =  72 \pi G n \omega^2 \frac{ (1 - n \cosh^2\omega t)}{\sinh^2 \omega t} = 72 \pi G n \omega^2\left[ \frac{1}{n\{1+(1+z)^{-\frac{2}{n}}\}}-1\right]
\end{equation}

We may calculate the flip time from eq.~\eqref{eq:40} as

\begin{equation}\label{eq:41}
t_{f} = \frac{1}{\omega} \cosh^{-1} \left(\sqrt{\frac{1}{n}} \right) = \frac{1}{\omega} cosh^{-1}\sqrt{\frac{3(1+\alpha)}{2}}
\end{equation}

which is identical with the  eq.~\eqref{eq:20}.

Also, from eq.~\eqref{eq:40}, we can derive the expression for the redshift at flip time $z_f$,
which is given by

\begin{equation}\label{eq:42}
z_f = \left(\frac{n}{1-n}\right)^{\frac{n}{2}} - 1
\end{equation}

The derived expression for the redshift at the flip time $z_f$ in eq.~\eqref{eq:42}
 matches  perfectly with earlier findings, as shown in eq.~\eqref{eq:28a}.

 The above consistencies in both approaches reinforce the robustness of the methodology and validate the theoretical framework.

\section{Comparative Analysis :}
In order to better understand the cosmological implications of the two formulations of the Generalized Chaplygin Gas (GCG) model discussed in this work, we perform a comparative analysis highlighting their key similarities and differences. The comparisons are made in terms of cosmological parameters, observational fits, and consistency with the standard $\Lambda$CDM model.

The following points summarize the main comparative features:

\begin{enumerate}[(i)]

\item One of the greatest achievements of the alternative approach is that it provides an explicit solution for the scale factor, demonstrating early deceleration followed by late-time acceleration. In contrast, the first approach was unable to obtain such a solution using the previous method.

\item We begin by discussing the Generalized Chaplygin Gas (GCG) model in terms of the deceleration parameter, $q$, at its extremal cases. The variation of $q$ with respect to the energy density $\rho$ and redshift $z$ is analyzed in detail.
\begin{itemize}
        \item[(a)] The GCG model depicts the evolution from a dust-dominated to an accelerating universe. For high $z$, it represents a dust-dominated phase, with $q = \frac{1}{2}$, matching the well-known $4D$ universe and indicating deceleration. At the present epoch ($ z = 0 $),  $q \approx -0.625$   suggests an accelerating universe. At the late stage of evolution, $ q = -1 $ corresponds to a pure $ \Lambda $CDM model.
        \item[(b)] In the alternative approach, the universe also appears dust-dominated at high $ z $. At $ z = 0 $, $ q \approx -0.247 $  signifies acceleration, and in later epochs, $ q = -1 $ leads to the $\Lambda$CDM model.
    \end{itemize}
Both approaches yield more or less similar conclusions.

In this context, it may not be inappropriate to mention that the primary advantage of the GCG model lies in its natural unification of dark matter and dark energy as different dynamical regimes of the same cosmic fluid. Its equation of state, which evolves from a pressureless, matter-like behavior at early times to a negative-pressure, dark energy-like state at late times, inherently addresses the coincidence problem without invoking separate, independently evolving components. This, in turn, reduces the need for fine-tuning multiple parameters to maintain consistency with observational data.

\item It is pertinent to mention that the flip time ($t_f$) has been derived using our alternative approach, as presented in eq.~\eqref{eq:20}, whereas it could not be obtained from the previous method. From fig.~\ref{tfa}, it is evident that the flip occurs at a later cosmic time for lower values of $\alpha$. Specifically, as the value of $\alpha$ decreases, $t_f$ correspondingly increases, indicating that a smaller $\alpha$ favors a later transition from deceleration to acceleration. Based on observations, we determine a small value of $\alpha = 0.0051$, as listed in Table~\ref{t20}.

\item
\begin{itemize}
        \item[(a)] Both eqs.~\eqref{eq:16a} and ~\eqref{eq:16b} define the redshift parameter at the flip time, $z_{f}$, which represents the point where the deceleration parameter changes sign. For the universe to be accelerating in the present epoch (	\textit{i.e.}, at $z = 0$), the condition $z_{f} > 0$ must be satisfied. From eq.~\eqref{eq:16a}, this requirement translates into the constraint $B > \frac{c}{2}$ or $\frac{B_s^{1+\alpha}}{\Omega_m} > \frac{1}{2}$, as mentioned in the first approach. This is also satisfied by the value obtained from our alternative approach, thereby confirming the consistency of the results. Furthermore, from eq.~\eqref{eq:16b}, we derive the constraint $\Omega_m < \frac{2}{3}$, which is consistent with current observational bounds on the matter density parameter. In our analysis, $\Omega_m = 0.2443$, a value well within the observational range. This agreement with observations supports the idea of an accelerating universe.

       \item[(b)] In our alternative approach, we also determine the redshift parameter $z_f$ at the flip time using eq.~\eqref{eq:28a}, which requires $n < 1$ for a real value of $z_f$. This condition aligns well with observational findings. For the universe to exhibit acceleration in the present epoch, we require $z_f > 0$, which further imposes the constraint $\alpha < \frac{1}{3}$. Specifically, for acceleration in the current epoch, we find that $\alpha < \frac{1}{3}$ (in our case, $\alpha = 0.0051$), a value that is consistent with those obtained from observational data.
 \end{itemize}

 \item
    \begin{itemize}
        \item[(a)] eq.~\eqref{eq:21}  provides the effective EoS in the GCG model. Initially, $ w_{\text{eff}} = 0 $ indicates a dust-dominated universe. At present epoch, $ w_{\text{eff}} \approx -0.75 $  implies acceleration. In the later epoch, $w_{\text{eff}} = -1 $ denotes the $\Lambda$CDM model.
        \item[(b)] In the alternative approach, eq.~\eqref{eq:18b}  describes the effective EoS ($w_e $). This expression requires $ \alpha < 1 $  for positive $z$ in the dust-dominated phase. Interestingly, for $\alpha = 1$ we find $ z = 0$, which seems to suggest that if we observe the universe  from dust dominated time scale, it appears as if the universe was filled with Chaplygin gas only. In the late universe,  $ w_e = -1 $ suggests the $\Lambda$CDM model, as discussed in section 3.2. At the present epoch ($ z = 0 $), $ w_e = - 0.9949 $ implies acceleration. For $\alpha = 1 $, $ w_e = 0 $, represents  a dust-dominated universe.
    \end{itemize}

\item A comparative study of the deceleration parameter and the effective equation of state for both approaches is presented in Table~\ref{tqw}.

\begin{table}[h!] \centering \begin{tabular}{|c|c|c|c|c|} \hline & \multicolumn{2}{c|}{1st approach} & \multicolumn{2}{c|}{2nd approach} \\ \hline $z$ & $q$ & $w_{\text{eff}}$ & $q$ & $w_e$ \\ \hline Dust dominated (High $z$)  & $0.5$ & $0$ & $0.5$ & $0$ \\ \hline Present epoch ($z=0$) & $-0.625$ & $-0.75$ & $-0.247$ & $-0.9949$ \\ \hline Late universe (negative $z$) & $-1$ & $-1$ & $-1$ & $-1$ \\ \hline \end{tabular} \caption{$q$ and effective EoS.} \label{tqw} \end{table}

\item  As calculated in eq.~\eqref{eq:18d} and illustrated in fig.~\ref{jz1}, the jerk parameter $j$ remains approximately unity throughout the entire cosmic evolution, with a slight increase above unity at the present epoch ($z = 0$). This small deviation reflects the influence of the GCG’s dark energy-like behavior beginning to emerge as the universe transits from a matter-dominated phase to an accelerated expansion phase. In contrast, our alternative formulation of the jerk parameter, given in eq.~\eqref{eq:29} and depicted in fig.~\ref{jza}, shows that $j$ slowly approaches unity at late cosmic times. This behavior effectively corresponds to a $\Lambda$CDM-like scenario at future epochs and remains consistent with current observational constraints.

  The Generalized Chaplygin Gas (GCG) model can act as a unifying fluid for both dark matter and dark energy, effectively reproducing the cosmic expansion history in close agreement with the standard $\Lambda$CDM model within observational limits. However, it offers a subtly different physical interpretation, potentially providing new insights into the nature of the universe’s accelerated expansion.

\item We have constrained the parameters using the Hubble $57$  dataset as follows:
\begin{itemize}
      \item[(a)] We obtain the values $\alpha = 0.03$ and $\Omega_m = 0.2443$ for the minimum $\chi^2$.
      \item[(b)] Using an alternative approach, we find $\Omega_m = 0.248$ and $\alpha = 0.0045$ from eq.~\eqref{eq:30}. Additionally, from eq.~\eqref{eq:31}, we obtain $B_s = 0.7532$ and $\alpha = 0.0051$.
\end{itemize}

In both approaches, the values of $\Omega_m$ are nearly identical. However, the value of $\alpha$ is slightly lower in the alternative approach.

\item The present age of the universe is also calculated for both approaches, yielding $13.95$ Gyr and
$13.52$ Gyr, and the results slightly differ from the age of $ 13.82$ Gyr obtained from the Planck 2020 data. This modest discrepancy arises primarily because the datasets probe distinct epochs and facets of cosmic evolution. Moreover, differences in the adopted values of the Hubble constant and other cosmological parameters across various analyses inevitably result in slight variations in the estimated age of the universe.

\end{enumerate}
In summary, both formulations of the GCG model successfully reproduce the essential features of the universe’s expansion history, exhibiting excellent agreement with the standard $\Lambda$CDM behavior at both early and late epochs.

\section{Discussion}

We have investigated the late-time acceleration of the universe within the framework of a spherically symmetric homogeneous model, employing a generalized Chaplygin gas (GCG) as the dark energy component. This study is motivated by the need to explain the observed acceleration of the universe's expansion. Two distinct approaches have been considered in this work. The governing equation, eq.~\eqref{eq:10}, is highly nonlinear and cannot be solved exactly in closed form. Previous studies addressed this by analyzing the system under extremal conditions — where the solution represents a dust-dominated universe at early times and asymptotically approaches the $\Lambda$CDM model for large values of the scale factor $a(t)$. However, this treatment does not allow for explicit prediction of the time evolution of the scale factor or the flip time marking the onset of acceleration.

To overcome this limitation, we proposed an alternative approximation by expanding the right-hand side of eq.~\eqref{eq:10} using a binomial series and retaining only the leading-order term. This approximation is justified in the late universe, where the scale factor is sufficiently large, and the ratio of model parameters remains small. Consequently, the truncated equation, presented in eq.~\eqref{eq:17}, admits an exact analytical solution for the scale factor, given in eq.~\eqref{eq:18}.

The generalization of the original Chaplygin gas model introduces an additional parameter $\alpha$, constrained within the range $0 < \alpha < \frac{1}{3}$ to ensure that the speed of sound remains subluminal, preserving causality. Furthermore, cosmological observations indicate that $\alpha$ should take small values in the late universe to achieve an optimal fit with empirical data.

 We present the best-fit curve of the redshift $z$ versus the Hubble parameter $H(z)$ in fig.~\ref{hz1}, obtained using the Hubble $57$ data set. Additionally, fig.~\ref{hz2} compares this best-fit curve with the theoretical prediction derived from eq.~\eqref{eq:21b}. The excellent agreement between these two curves across the entire range of cosmic evolution demonstrates the consistency of our model with observational data. Furthermore, we analyze the behavior of the deceleration parameter, the effective equation of state, the jerk parameter, and other key cosmological quantities as functions of redshift $z$.

Subsequently, the entire analysis has been reinterpreted within the framework of the Raychaudhuri equation. As expected, the outcomes are broadly consistent with previous studies, thereby reinforcing the validity of our approximation and approach. It is worth noting that while the GCG model has been extensively explored over the past two decades, most analyses have focused primarily on its limiting cases. In contrast, the present study offers an exact solution for the scale factor within the proposed approximation, allowing a detailed investigation of the universe's expansion history, including the explicit determination of the flip time. Moreover, the low values of $\alpha$ obtained through both the Hubble $57$ data and alternative methods lend further support to the model's reliability in accurately capturing the dynamics of late-time cosmic acceleration.

\textbf{Acknowledgments}

 \vspace{0.1 cm}
 DP acknowledges Dr. S. Chatterjee for valuable comments and suggestions. DP also acknowledges the financial support of Netaji Nagar Day College for a Minor Research Project.


\vspace{0.2 cm}
\begin{thebibliography}{15}
\bibitem{res} Reiss \textit{et. al.}, \textit{Astro. Phys. Jour. }\textbf{607} 665(2004); astro-ph/9805201.

\bibitem{spe} D. N. Spergel \textit{et. al.},  \textit{Astro. Phys. Jour. Suppl.} \textbf{148} 175 (2003).

\bibitem{cop} E. Copeland E, M. Sami and S. Tsujikawa, \textit{Int Jour. Mod. Phys.} \textbf{D15} 1753(2006).

\bibitem{sam} M. Sami and T. Padmanabhan, \textit{ Phys. Rev.} \textbf{D67} 083509 (2003).

\bibitem{sch} R. J. Scherrer, \textit{ Phys. Rev. Lett.} \textbf{93} 011301 (2004).

\bibitem{gib} G. W. Gibbons, \textit{Phys. Lett.} \textbf{B 537} 1 (2002).

\bibitem{eli} E. Elizalde, S. Nojiri and S. Odintsov,  \textit{Phy. Rev. } \textbf{D70} 043543 (2004).

\bibitem{guo} Z. Guo, Y. Piao, X. Zhang and Y. Zhang,  \textit{Phy. Rev.} \textbf{D74} 127304 (2006).

\bibitem{wan} M. Wanas, `Dark Energy: Is It of Torsion Origin?',  Proceedings of the first MEARIM,
    edited by A. A. Hady and M. I. Wannas, P-41 (2009); arXiv:1006.2154v1[gr-qc].


\bibitem{neu} I. P. Neupane,  \textit{Class. Quant. Grav.} \textbf{26} 195008 (2009); arXiv:0905.2774[hep-th];
          I. P. Neupane,  \textit{Int. J. Mod. Physics} \textbf{D19} 2281 (2010); arXiv:1004.0254v1[gr-qc].

\bibitem{dp1} D. Panigrahi and S. Chatterjee, \textit{ Grav. Cosm.} \textbf{17} 18 (2011); gr-qc/1006.0476.

 \bibitem{dp2}   D. Panigrahi, S. Chatterjee and Y. Z. Zhang, \textit{Int. Jour. Mod. Phys.} \textbf{A21} 6491 (2006); gr-qc/0604079.

\bibitem{sah} Varun Sahni amd Yuri Shtanov, `Cosmic Acceleration and Extra    Dimensions' arXiv:0811.3839v1 [astro-ph]; S. Kachru, R. Kallosh R. Linde and S. P. Trivedi, \textit{Phys. Rev. } \textbf{D68} 046005 (2003);  M. S. Carroll and L. Mersini, \textit{Phys. Rev.}  \textbf{D64} 124008 (2001).

\bibitem{kra}Andrzej Krasinski, Charles Hellaby, Krzysztof Bolejko and Marie-Noelle Celerier, \textit{ Gen. Rel. Grav.} \textbf{42} 2453 (2010) arXiv: 0903.4070v2; H. Alnes, A. Morad  and $\O~$ Gron,   \textit{J. Cosmol. Astropart. Phys.} \textbf{01} 007 (2007); S. Chatterjee, \textit{J. Cosmol. Astropart. Phys.} \textbf{03} 014 (2011); C. M. Hirata and U. Seljak,  \textit{Phys. Rev.}  \textbf{D72} 083501 (2005) ; astro-ph/0503582.

\bibitem{bent} M. C. Bento, O. Bertolami and A. A. Sen,  \textit{Phys. Rev.} \textbf{D66} 043507 (2002); V. Gorini, A. Kamenschik and U. Moschella,   \textit{Phys. Rev.} \textbf{D67} 063509 (2003).

\bibitem{bento} M. C. Bento, O. Bertolami and A. A. Sen,  \textit{Phys.Rev.}\textbf{D67}  063003 (2003).

\bibitem{bert} O. Bertolami, A.A.Sen, S. Sen, P.T. Silva, \textit{ Mon. Not. Roy. Astron. Soc.} \textbf{353} 329 (2004);astro-ph/0402387


\bibitem{wu} P. Wu and H.Yu, \textit{Phys. Lett}\textbf{B 644} 16 (2007); hep-th/9307036.

\bibitem{thak} P. Thakur, \textit{Pramana J. Phys.}\textbf{88} 51 (2017); hep-th/9307036.


\bibitem{bord} M. Bordemann and J. Hoppe,   \textit{Phys. Lett}\textbf{B 317} 315 (1993); hep-th/9307036.

\bibitem{hop} J. Hoppe, (1993);  hep-th/9311059.

\bibitem{jac} R. Jackiw and A. P. Polychronakos, \textit{Phys. Rev.}\textbf{D62} 085019 (2000); hep-th/0004083.

\bibitem{zla} I. Zlatev, L. M. Wand and P. J. Steinhardt,  \textit{Phys. Rev. Lett}\textbf{82} 896 (1999); astro-ph/9807002.


\bibitem{blan} R. D. Blandford et al., \textit{ASP Conf. Ser.} \textbf{339} 27 (2004); astro-ph/0408279.


\bibitem{rap}D. Rapetti, S. W. Allen, M. A. Amin and R. D. Blandford, \textit{ Mon. Not. Roy. Astron. Soc.} \textbf{375} 1510 (2007).



\bibitem{sah1}V. Sahni, T. D. Saini, A. A. Starobinsky, U. Alam, \textit{JETP Lett.} \textbf{77} 201 (2003).


\bibitem{alam} U. Alam, V. Sahni, T. D. Saini, A. A. Starobinsky, \textit{ Mon. Not. Roy. Astron. Soc.} \textbf{344} 1057 (2003).

\bibitem{sha} G. S. Sharov and V. O. Vasiliev, \textit{Math.  Modelling and Geo}. \textbf{6} 1 (2018).

\bibitem{zhang} C. Zhang \emph{et al.}  \textit{RAA (ZResearch in Astronomy and Astrophysics)} \textbf{14} 1221 (2014).

\bibitem{ster} D. Stern, R. Jimenez, L. Verde, M. Kamionkowski and S. A. Stanford,  \textit{J. Cosmol. Astropart. Phys.} \textbf{02} 008 (2010).

\bibitem{mor1} M. Moresco, L. Verde, L. Pozzetti, R. Jimenez and A. Cimatti, \textit{J. Cosmol. Astropart. Phys.} \textbf{7}  053 (2012).

\bibitem{mor2} M. Moresco, L. Pozzetti, A. Cimatti, R. Jimenez, C. Maraston, L. Verde, D. Thomas, A. Citro, R. Tojeiro and D. Wilkinson, \textit{J. Cosmol. Astropart. Phys.} \textbf{05} 014 (2016).


\bibitem{rats} A. L. Ratsimbazafy, S. I. Loubser, S. M. Crawford, C. M. Cress, B. A. Bassett, R. C. Nichol and R. Visnen,\textit{ Mon. Not. Roy. Astron. Soc.} \textbf{467} 3239 (2017).

\bibitem{mor3} M. Moresco, \textit{ Mon. Not. Roy. Astron. Soc.} \textbf{450}, L16 (2015).

\bibitem{gaz} E. Gaztanaga, A. Cabre and L. Hui, \textit{ Mon. Not. Roy. Astron. Soc.} \textbf{399} 1663 (2009).


\bibitem{oka} A. Oka  \textit{et al.}, \textit{ Mon. Not. Roy. Astron. Soc.} \textbf{ 439}  2515(2014); arXiv: 1310.2820.

\bibitem{wan1}  Y. Wang  \textit{et al.},  \textit{ Mon. Not. Roy. Astron. Soc.}\textbf{469} 3762 (2017); arXiv: 1607.03154

\bibitem{chu} C.H.  Chuang  and Y. Wang,   \textit{ Mon. Not. Roy. Astron. Soc.} \textbf{435} 255 (2013); arXiv: 1209.0210.


\bibitem{ala} S. Alam \textit{ et al. }, \textit{ Mon. Not. Roy. Astron. Soc.} \textbf{ 470} 2617 (2017); arXiv: 1607.0315.)

\bibitem{bla} C. Blake \textit{et al.},  \textit{ Mon. Not. Roy. Astron. Soc.}  \textbf{425} 405 (2012; arXiv: 1204.3674.

\bibitem{del} T. Delubac, J. Rich, S. Bailey  \emph{et al.}, \textit{Astronomy and Astrophysics} \textbf{A96}, 552 (2013).


\bibitem{bus}  N.G. Busca  \textit{et al.},  \textit{Astron. and Astrop.} \textbf{552} A96 (2013); arXiv: 1211.2616.

\bibitem{bau} J. E. Bautista  \textit{et al.},  \textit{Astron. Astrophys.}  \textbf{603} A12 (2017); arXiv: 1702.00176.

\bibitem{fon} A.  Font-Ribera \textit{ et al.}, \textit{J. Cosmol. Astropart. Phys.} \textbf{ 05} 027 (2014); arXiv: 1311.1767.

\bibitem{sei} M. Seikel, S. Yahya, R. Maartens and C. Clarkson, \textit{Phys. Rev.} \textbf{D 86} 083001 (2012).

\bibitem{seth}  G. Sethi, S. K. Singh and P. Kumar, \textit{Int. J. Mod. Phys.} \textbf{D15} 1089 (2006).

\bibitem{male} M. Malekjani, A. Khodam-Mohammadi and N. Nazari-pooya, \textit{Astrophys Space Sci} \textbf{334} 193 (2011).

\bibitem{pla} E. Rosenberg, S. Gratton, G. Efstathiou,  \textit{ Mon. Not. Roy. Astron. Soc.}  \textbf{517} 4620 (2022); arXiv: 2205.10869.

\bibitem{na} N. Aghanim, Y. Akrami \textit{et al.},  \textit{Astron. and Astrop.} \textbf{641} A6 (2020).

\bibitem{dp3} D. Panigrahi,  IOP Conf. Series: Journal of Physics: Conf. Series \textbf{1251} 012039(2019).

\bibitem{dp4} D. Panigrahi, B. C. Paul and S. Chatterjee, \textit{Eur. Phys. J. Plus} \textbf{136} 771(2021); arxiv: 2104.12169[gr-qc].


\bibitem{ray}  A. K. Raychaudhuri, \textit{Phys. Rev.} \textbf{98}, 1123 (1955).




\end{thebibliography}
\end{document}